\documentclass[a4paper,11pt]{article}

\usepackage[numbers,sort&compress]{natbib}
\usepackage[all]{xy}
\usepackage{amsmath,amsfonts,amssymb,amsthm,epsfig,amscd,comment,latexsym,psfrag}
\usepackage{CJK}
\usepackage{mathrsfs}
\usepackage{nicefrac,xspace,tikz}
\usepackage{arydshln}
\usetikzlibrary{arrows}
\usepackage{xcolor,float}
\usepackage{appendix}
\usepackage{txfonts}
\usepackage{verbatim}
\usepackage{multirow}
\makeatletter

\newcommand{\Rmnum}[1]{\expandafter\@slowromancap\romannumeral #1@}
\makeatother
\def\footnoterule{\kern 1mm \hrule width 7cm \kern 2.2mm}%

\def\dprod{\displaystyle\prod}

\def\dsum{\displaystyle\sum}

\def\Tr{\mathrm{Tr}}

\numberwithin{equation}{section}

\renewcommand{\a}{\alpha}
\renewcommand{\b}{\beta}

\usepackage{graphicx}
\usepackage{float}
\usepackage{subfigure}
\usepackage[]{caption2}

\renewcommand{\thesubfigure}{(\roman{subfigure})}
\makeatletter \renewcommand{\@thesubfigure}{\thesubfigure \space}
\renewcommand{\p@subfigure}{} \makeatother

\renewcommand\appendix{\par
    \setcounter{section}{0}
    \setcounter{subsection}{0}
    \gdef\thesection{Appendix \Alph{section}}}

\allowdisplaybreaks
\begin{document}
\begin{titlepage}
\begin{center}
{\Large\bf
Correlators in two rainbow tensor and complex multi-matrix models}\vskip .2in
{\large Bei Kang$^{a}$,\footnote{kangbei@ncwu.edu.cn}
Lu-Yao Wang$^{b,c}$,\footnote{wangluyao@bimsa.cn}  Ke Wu$^{d}$\footnote{wuke@cnu.edu.cn}
and Wei-Zhong Zhao$^{d,}$\footnote{Corresponding author: zhaowz@cnu.edu.cn}} \vskip .2in
$^a${\em School of Mathematics and Statistics, North China University of Water Resources and Electric Power,
Zhengzhou 450046, Henan, China}\\
$^b${\em Beijing Institute of Mathematical Sciences and Applications, Beijing, 101408, China}\\
$^c${\em Yau Mathematical Sciences Center, Tsinghua University, Beijing, 100084, China} \\
$^d${\em School of Mathematical Sciences, Capital Normal University,
Beijing 100048, China}

\begin{abstract}
We construct two rainbow tensor models with multi-tensors of rank-$3$ and present their $W$-representations.
We give the formula of counting number of independent gauge-invariant operators in terms of Hurwitz numbers
and establish a one-to-one correspondence between connected operators and colored Dessins.
By means of the colored Dessins and $W$-representations, respectively, we derive two compact expressions of correlators
for each of rainbow tensor models. Furthermore, two complex multi-matrix models from the degradations of the
constructed rainbow tensor models are also discussed.

\end{abstract}

\end{center}


\end{titlepage}

\section{Introduction}
Tensor models are generalizations of matrix models from matrices to tensor. Since they provide the analytical tool for the study of
random geometries in three and more dimensions, much interest has been attributed to their remarkable properties.
The counting of tensor model invariants has been extensively studied \cite{jbgeloun1}-\cite{RAvohou}. Ben Geloun and Ramgoolam  \cite{jbgeloun1}
counted gauge invariants for tensor models, which exhibit a relationship with counting problems of branched covers of the 2-sphere.
The explicit generating functions for the relevant counting and some formulae for correlators of the tensor model invariants were presented.
$U(N)^{\bigotimes r}\bigotimes O(N)^{\bigotimes q}$ invariants are tensor model observables endowed with a tensor field of order $(r, q)$.
These observables were recently enumerated by using group theoretic formulae \cite{RAvohou}. The Grothendieck's Dessin d'Enfant,
which is referred as Dessin for short in this paper, is a bipartite graph embedded into an orientable two-dimensional surface.
Itoyama et al. \cite{Amburg,feydessins} gave the operator/Feynman diagram/Dessin correspondence in the tensor model with a tensor of rank-$r$.
This correspondence indicates a classification of operators in the rank-$r$ tensor models by Feynman diagrams in the rank-$(r-1)$ tensor models.
When particularized to the case of rank two-rank three correspondence, it can be combined with the geometrical pictures of the dual of the original
Feynman diagrams, namely, equilateral triangulations (Dessins) to form a triality. Based on the operator/Dessin correspondence,
the cut and join operations were given in the language of Dessins.

Rainbow tensor models have attracted considerable attention \cite{ItoyamaJHEP2017}-\cite{ItoyamaJHEP2019}. Itoyama et al. \cite{ItoyamaJHEP2017}
analyzed a rainbow tensor model, i.e., so-called the Aristotelian RGB (red-green-blue) model and introduced a few methods which allow one to connect
calculations in the tensor models to those in the matrix models. $W$-representation of matrix model gives the dual expression for the partition function
through differentiation rather than integration \cite{Morozov09}-\cite{MirGKM}. It makes the calculation of the correlators and analysis of superintegrability
for matrix models very convenient \cite{Mironov2022PLB}-\cite{LWYZ}. Attempts have also been made for the case of tensor models \cite{Itoyama2020}-\cite{BKang2024}.
Recently, the $W$-representations of a rainbow tensor model with one tensor \cite{BKang2021} and a two-tensor model with order-three \cite{BKang2024} were derived.
With the help of $W$-representations, the correlators of these two models were computed exactly.

Hurwitz numbers count branched coverings of the sphere by a Riemann surface with prescribed ramification profiles. The generating function of Hurwitz numbers
can be represented in terms of the cut-and-join operator. The exponential of the generating function for double Hurwitz numbers gives a tau-function of the KP
hierarchy \cite{Shcherbin}-\cite{Goulden}. Ambj$\text{\o{}}$rn and Chekhov \cite{Leonid} proved that generating functions for numbers of three different
types of Belyi morphisms are free energies of special matrix models all of which are tau functions of the KP hierarchy. For the Hermitian one-Matrix model with
a general potential, Koch and Ramgoolam \cite{Galois} showed that the correlators can be mapped to the counting of certain triples of permutations.
Moreover, correlators of multi-matrix models can be mapped to the counting of triples of permutations subject to equivalences defined by subgroups of the permutation
groups. This was related to colorings of the edges of the Grothendieck Dessins \cite{Galois}.

In this paper, we shall construct the rainbow tensor models with multi-tensors of rank-$3$, which can be realized by $W$-representations. Thus the correlators can be
calculated by means of $W$-operators. To provide a new way of calculation of correlators, a one-to-one correspondence between connected operators and colored
Dessins will be established. An immediate benefit of this correspondence is the possibility of calculation of correlators from the viewpoint of colored Dessins.

This paper is organized as follows. In Sect.2, we analyze the gauge-invariant operators which are constructed by contractions of all indices of tensors
and their complex conjugates. Then we give the formula of counting number of the independent operators. In Sect. 3, we establish a one-to-one correspondence between
connected operators and colored Dessins. For the disconnected operator consisting of $i$ connected operators, it corresponds to $i$-colored Dessins.
In Sect. 4, we first construct a rainbow tensor model with multi-tensors of rank-$3$ in terms of the connected operators, which can be realized by the
$W$-representation. We give two compact expressions of correlators by means of the colored Dessins and $W$-representations, respectively. As an example,
we consider the rainbow tensor model with two tensors and calculate the correlators in detail from the viewpoint of colored Dessins. Then we construct
another rainbow tensor model with $W$-representation and present also two compact expressions of correlators.
In Sect. 5, we give two complex multi-matrix models from the degradation of the rainbow tensor models and discuss their correlators. We end this paper with
the conclusion in Sect. 6.


\section{Counting gauge-invariant operators }

In rainbow tensor models, there are $\sharp_L$ complex tensors $T_{i_{1}, \cdots, i_{r}}^{j}$
with the ``gauge'' symmetry $\mathcal{U} = U(N_1)\otimes \cdots \otimes U(N_r)$,
where $j \in \{{1, 2,\cdots, \sharp_L}\}$, $\sharp_L$ denotes the number of tensors and $r$ is called the rank of the
fields $T_{i_{1}, \cdots, i_{r}}^{j}$\cite{ItoyamaNPB2018}.
In this paper, we focus on tensors of rank-$3$.

Note that the operators of interest are invariants of $\mathcal{U}$.
They are made by contractions of all indices of tensors $T_{i_{1} ,i_2, i_{3}}^{j}$
and their complex conjugates $\bar{T}^{i_{1},i_{2}, i_{3}}_{j}$.
 We denote invariants formally as
\begin{eqnarray}\label{rsigma}
\mathcal{R}_{\tilde{\sigma}}^{(\vec{a},\vec{b})}&=&
\mathcal{R}_{(\sigma_1,\sigma_2,\sigma_3)}^{(\vec{a},\vec{b})}\nonumber\\
&=&\mathcal{M}_{I_1,I_2,I_3}\cdot(\sigma_1,\sigma_2,\sigma_3)\cdot\bar{\mathcal{M}}^{I_1,I_2,I_3}\nonumber\\\
&=&\mathcal{M}_{I_1,I_2,I_3}\bar{\mathcal{M}}^{\sigma_1(I_1),\sigma_2(I_2),\sigma_3(I_3)},
\end{eqnarray}
where~$(\vec{a},\vec{b}):=(a_1,\cdots, a_{\sharp_L}, b_1,\cdots,b_{\sharp_L})$, $a_i$ and $b_i$
are non-negative integers satisfying $\sum_{i=1}^{\sharp_L}a_i=\sum_{i=1}^{\sharp_L}b_i\geqslant1$.
Here we define $\sum_{i=1}^{\sharp_L}a_i$ or $\sum_{i=1}^{\sharp_L}b_i$ as the level of
$\mathcal{R}_{\tilde{\sigma}}^{(\vec{a},\vec{b})}$,
and denote $level (\mathcal{R}_{\tilde{\sigma}}^{(\vec{a},\vec{b})})=\sum_{i=1}^{\sharp_L}a_i=\sum_{i=1}^{\sharp_L}b_i$.
The element~$\tilde{\sigma}:=(\sigma_1,\sigma_2,\sigma_3)\in
S_{\sum_{i=1}^{\sharp_L}a_i}\times S_{\sum_{i=1}^{\sharp_L}a_i}\times S_{\sum_{i=1}^{\sharp_L}a_i}$
acts independently on three indices of~$\bar{\mathcal{M}}^{I_1,I_2,I_3}$,
where~$S_{\sum_{i=1}^{\sharp_L}a_i}$ is the permutation group.
The sleek notation uses the capital Roman letters to collect all of the little Roman letter indices,
i.e. $I_j$ stands for $i_j^{(1)},i_j^{(2)},\cdots,i_j^{(a_1+\cdots+a_{\sharp_L})},j=1,2,3$.
There is always an implicit summation whenever a pair of repeated up-down indices appeared in this paper.

$\mathcal{M}_{I_1,I_2,I_3}$ and $\bar{\mathcal{M}}^{\sigma_1(I_1),\sigma_2(I_2),\sigma_3(I_3)}$ in (\ref{rsigma})
are given by
\begin{eqnarray}\label{Mfield1}
\mathcal{M}_{I_1,I_2,I_3}&=&T^1_{i_1^{(1)},i^{(1)}_2,i^{(1)}_{3}}\cdots T^1_{i_1^{(a_1)},i^{(a_1)}_2,i^{(a_1)}_3}\cdots
T^{\sharp_L}_{i_1^{(a_1+\cdots+a_{\sharp_L-1}+1)},i^{(a_1+\cdots+a_{\sharp_L-1}+1)}_2,i^{(a_1+\cdots+a_{\sharp_L-1}+1)}_3}\nonumber\\
&&\cdots T^{\sharp_L}_{i_1^{(a_1+\cdots+a_{\sharp_L})},i^{(a_1+\cdots+a_{\sharp_L})}_2,i^{(a_1+\cdots+a_{\sharp_L})}_3},
\end{eqnarray}
\begin{eqnarray}\label{Mfield2}
\bar{\mathcal{M}}^{I_1,I_2,I_3}&=&\bar{T}_1^{i_1^{(1)},i^{(1)}_2,i^{(1)}_{3}}\cdots \bar{T}_1^{i_1^{(b_1)},i^{(b_1)}_2,i^{(b_1)}_3}
\cdots \bar{T}_{\sharp_L}^{i_1^{(b_1+\cdots+b_{\sharp_L-1}+1)},i^{(b_1+\cdots+b_{\sharp_L-1}+1)}_2,i^{(b_1+\cdots+b_{\sharp_L-1}+1)}_3}\nonumber\\
&&\cdots \bar{T}_{\sharp_L}^{i_1^{(b_1+\cdots+b_{\sharp_L})},i^{(b_1+\cdots+b_{\sharp_L})}_2,i^{(b_1+\cdots+b_{\sharp_L})}_3},
\end{eqnarray}
\begin{eqnarray}\label{barmsigma}
&&\bar{\mathcal{M}}^{\sigma_1(I_1),\sigma_2(I_2),\sigma_3(I_3)}=(\sigma_1,\sigma_2,\sigma_3)\cdot\bar{\mathcal{M}}^{I_1,I_2,I_3}
=\bar{T}_1^{i_1^{\sigma_1(1)},i^{\sigma_2(1)}_2,i^{\sigma_3(1)}_3}\cdots
\bar{T}_1^{i_1^{\sigma_1(b_1)},i^{\sigma_2(b_1)}_2,i^{\sigma_3(b_1)}_3}\nonumber\\
&&
\cdots\bar{T}_{\sharp_L}^{i_1^{\sigma_1(b_1+\cdots+b_{\sharp_L-1}+1)},
i^{\sigma_2(b_1+\cdots+b_{\sharp_L-1}+1)}_2,i^{\sigma_3(b_1+\cdots+b_{\sharp_L-1}+1)}_3}\cdots
\bar{T}_{\sharp_L}^{i_1^{\sigma_1(b_1+\cdots+b_{\sharp_L})},i_2^{\sigma_2(b_1+\cdots+b_{\sharp_L})},i^{\sigma_3(b_1+\cdots+b_{\sharp_L})}_3}.
\end{eqnarray}
Here we denote~$\sigma_k(I)=(i^{\sigma_k(1)},\cdots,i^{\sigma_k(b_1+\cdots+b_{\sharp_L})}),k=1,2,3$.
We see that operators~$\mathcal{R}_{\tilde{\sigma}}^{(a_1,0,\cdots,0,b_1,0,\cdots,0)}$, $\cdots,$
$\mathcal{R}_{\tilde{\sigma}}^{(0,\cdots,0,a_{\sharp_L},0\cdots,0,b_{\sharp_L})}$ are
exactly the gauge invariants in the Aristotelian RGB model \cite{ItoyamaJHEP2017}.

Let us count the number of invariants~$\mathcal{R}_{\tilde{\sigma}}^{(\vec{a},\vec{b})}$.
It depends on the number of zeros for $a_i$ and $b_i$, $i=1,\cdots,\sharp_L$.

(i) There is only one positive integer for $a_i$ and $b_i$, respectively.

Let us take $a_1$, $b_2$ to be nonzero without loss of generality.
This case is similar as~$\mathcal{R}_{\tilde{\sigma}}^{(a_1,0,\cdots,0,b_1,0,\cdots,0)}$.
Permutations~$\sigma_i$, $i=1,2,3$, in~$\mathcal{R}_{\tilde{\sigma}}^{(a_1,0,\cdots,0,b_2,0,\cdots,0)}$
are characterized by the double coset~$S_{a_1}\setminus S_{a_1}\times S_{a_1}\times S_{a_1}/S_{a_1}$.
The number of elements in this double coset is
$|S_{a_1}\setminus S_{a_1}\times S_{a_1}\times S_{a_1}/S_{a_1}|=\sum_{\zeta\vdash a_1}Sym(\zeta)$,
where the sum over~$\zeta=(h_1,h_2,\cdots,h_{a_1})$ is performed over all partitions of~$a_1=\sum_iih_i$,
and~$h_i$ gives the number of cycles of size~$i$ in~$\zeta$.
$Sym(\zeta)=\prod_{i=1}^{a_1}(i^{h_i})(h_i!)$
is the number of elements of~$S_{a_1}$ commuting with any permutation in the conjugacy class $T_\zeta$ determined by~$\zeta$.

(ii) There is only one nonzero integer for $a_i$ or $b_i$.

Let us take $a_1,b_1,\cdots,b_l,2\leqslant l\leqslant \sharp_L$ to be nonzero.
Thus we have~$\sum_{i=1}^lb_i=a_1$. Define $H_{b_1,\cdots,b_l}$ as a subgroup of~$S_{b_1+\cdots+b_l},$
in which the permutations map $b_1,\cdots,b_l$ elements to $b_1,\cdots,b_l$ elements, successively.
For~$\alpha_1\in S_{a_1}$ and~$\alpha_2\in H_{b_1,\cdots,b_l}$, we have
\begin{eqnarray}\label{}
\mathcal{M}_{I_1,I_2,I_3}\cdot(\sigma_1,\sigma_2,\sigma_3)\cdot\bar{\mathcal{M}}^{I_1,I_2,I_3}&=&
(\mathcal{M}_{I_1,I_2,I_3}\cdot \alpha_1)\cdot(\sigma_1,\sigma_2,\sigma_3)\cdot(\alpha_2 \cdot\bar{\mathcal{M}}^{I_1,I_2,I_3})\nonumber\\
&=&\mathcal{R}_{(\alpha_1\sigma_1\alpha_2,\alpha_1\sigma_2\alpha_2,\alpha_1\sigma_3\alpha_2)}^{(a_1,0,\cdots,0,b_1,\cdots,b_l,0,\cdots,0)}.
\end{eqnarray}
Hence $\mathcal{R}_{\tilde{\sigma}}^{(a_1,0,\cdots,0,b_1,\cdots,b_l,0,\cdots,0)}$ and
$\mathcal{R}_{(\alpha_1\sigma_1\alpha_2,\alpha_1\sigma_2\alpha_2,\alpha_1\sigma_3\alpha_2)}^{(a_1,0,\cdots,0,b_1,\cdots,b_l,0,\cdots,0)}$
are indeed the same gauge invariant operator. It implies that~$\sigma_i$, $i=1,2,3$,
in~$\mathcal{R}_{\tilde{\sigma}}^{(a_1,b_1,\cdots,b_l)}$ are characterized by the double coset
$S_{a_1}\setminus S_{a_1}\times S_{a_1}\times S_{a_1}/H_{b_1,\cdots,b_l}$ \cite{Tribelhorn,jbgeloun1,jbgeloun2}.

For a permutation $\rho$ in~$S_n$, the size of conjugacy class $|\mathcal{T}_\rho|$ is given by $|\mathcal{T}_\rho|=\frac{n!}{Sym(\rho)}$.
By Burnside's Lemma, we have
\begin{eqnarray}\label{nedc1}
&&|S_{a_1}\setminus S_{a_1}\times S_{a_1}\times S_{a_1}/H_{b_1,\cdots,b_l}|\nonumber\\
&=&\frac{1}{|S_{a_1}||H_{b_1,\cdots,b_l}|}\sum_{\alpha_2\in H_{b_1,\cdots,b_l}}
\sum_{\alpha_1,\sigma_1,\sigma_2,\sigma_3\in S_{a_1}}\delta(\alpha_1\sigma_1\alpha_2\sigma_1^{-1})
\delta(\alpha_1\sigma_2\alpha_2\sigma_2^{-1})\delta(\alpha_1\sigma_3\alpha_2\sigma_3^{-1})\nonumber\\
&=&\frac{1}{a_1!|H_{b_1,\cdots,b_l}|}\sum_{\alpha_2\in H_{b_1,\cdots,b_l}}
\frac{a_1!}{Sym(\alpha_2)}(Sym(\alpha_2))^3\nonumber\\
&=&\frac{1}{|H_{b_1,\cdots,b_l}|}\sum_{\zeta\in H_{b_1,\cdots,b_l}}(Sym(\zeta))^2,
\end{eqnarray}
where $\delta$ is the Kronecker delta of the symmetric group defined to be equal to one when the argument is the identity and zero otherwise.
For the given~$\alpha_2$ and~$\alpha_1$ in $\mathcal{T}_{\alpha_2}$ in the second line of (\ref{nedc1}),
$\delta(\alpha_1\sigma_i\alpha_2\sigma_i^{-1}),i=1,2,3$, count the permutations in~$ S_{a_1}$ commutating with any permutation in the conjugacy class of~$\alpha_2$,
which is exactly~$Sym(\alpha_2)$. As the fact that the size of the conjugacy class~$\mathcal{T}_{\alpha_2}$ in~$S_{a_1}$ is $\frac{a_1!}{Sym(\alpha_2)}$,
we obtain the third line of (\ref{nedc1}).

(iii) There are more than one positive integers for $a_i$ and $b_i$, respectively.

Let us take $a_1,\cdots,a_{l_1},b_1,\cdots,b_{l_2},2\leqslant l_1,l_2\leqslant \sharp_L$ to be nonzero.
We define $H_{a_1,\cdots,a_{l_1}}$ and $H_{b_1,\cdots,b_{l_2}}$ as subgroups of~$S_{a_1+\cdots+a_{l_1}}$,
in which the permutations map $a_1,\cdots,a_{l_1}$ elements to $a_1,\cdots,a_{l_1}$ elements and
$b_1,\cdots,b_{l_2}$ elements to $b_1,\cdots,b_{l_2}$ elements, respectively.

Similarly, it can be shown that
$\mathcal{R}_{\tilde{\sigma}}^{(a_1,\cdots,a_{l_1},0,\cdots,0,b_1,\cdots,b_{l_2},0,\cdots,0)}$
and $\mathcal{R}_{(\alpha_1\sigma_1\alpha_2,\alpha_1\sigma_2\alpha_2,\alpha_1\sigma_3\alpha_2)}
^{(a_1,\cdots,a_{l_1},0,\cdots,0,b_1,\cdots,b_{l_2},0,\cdots,0)}$
for~$\alpha_1\in H_{a_1,\cdots,a_{l_1}}$ and~$\alpha_2\in H_{b_1,\cdots,b_{l_2}}$
are the same gauge invariant operator.
Thus~$\sigma_i$, $i=1,2,3$, in~the operator
$\mathcal{R}_{\tilde{\sigma}}^{(a_1,\cdots,a_{l_1},0,\cdots,0,b_1,\cdots,b_{l_2},0,\cdots,0)}$
are characterized by the double coset
$H_{a_1,\cdots,a_{l_1}}\setminus S_{a_1+\cdots+a_{l_1}}\times
S_{a_1+\cdots+a_{l_1}}\times S_{a_1+\cdots+a_{l_1}}/H_{b_1,\cdots,b_{l_2}}$.

By Burnside's Lemma, the number of elements in this double coset is
\begin{eqnarray}\label{nedc2}
&&|H_{a_1,\cdots,a_{l_1}}\setminus S_{a_1+\cdots+a_{l_1}}\times S_{a_1+\cdots+a_{l_1}}
\times S_{a_1+\cdots+a_{l_1}}/H_{b_1,\cdots,b_{l_2}}|
=\frac{1}{|H_{a_1,\cdots,a_{l_1}}||H_{b_1,\cdots,b_{l_2}}|}
\nonumber\\
&&\times \sum_{\alpha_1\in H_{a_1,\cdots,a_{l_1}}}
\sum_{\alpha_2\in H_{b_1,\cdots,b_{l_2}}}\sum_{\sigma_1,\sigma_2,\sigma_3\in S_{a_1+\cdots+a_{l_1}}}
\delta(\alpha_1\sigma_1\alpha_2\sigma_1^{-1})\delta(\alpha_1\sigma_2\alpha_2\sigma_2^{-1})\delta(\alpha_1\sigma_3\alpha_2\sigma_3^{-1})
\nonumber\\
&&=\frac{(a_1+\cdots+a_{l_1})!}{|H_{a_1,\cdots,a_{l_1}}||H_{b_1,\cdots,b_{l_2}}|}
\sum_{\zeta\in H_{a_1,\cdots,a_{l_1}}\cap H_{b_1,\cdots,b_{l_2}}}(Sym(\zeta))^2.
\end{eqnarray}

For the independent operators at each level-$l$, they consist of above three cases. Hence their number is
\begin{eqnarray}\label{openum}
\tilde{\sharp}_{l}&=&\sum_{l_1=2}^{min\{\sharp_L,l\}}\binom{\sharp_L}{l_1}
\sum_{\substack{b_1+\cdots+b_{l_1}=l\\b_1,\cdots,b_{l_1}\geqslant 1}}
\frac{2\sharp_L}{|H_{b_1,\cdots,b_{l_1}}|}\sum_{\zeta\in H_{b_1,\cdots,b_{l_1}}}
(Sym(\zeta))^2+\sharp_L^2\sum_{\zeta\vdash l}Sym(\zeta)\nonumber\\
&&+\sum_{l_1,l_2=2}^{min\{\sharp_L,l\}}\binom{\sharp_L}{l_1}\binom{\sharp_L}{l_2}
\sum_{\substack{a_1+\cdots+a_{l_1}=l\\a_1,\cdots,a_{l_1}\geqslant 1}}
\sum_{\substack{b_1+\cdots+b_{l_2}=l\\b_1,\cdots,b_{l_2}\geqslant 1}}
\frac{l!}{|H_{a_1,\cdots,a_{l_1}}||H_{b_1,\cdots,b_{l_2}}|}
\sum_{\zeta\in H_{a_1,\cdots,a_{l_1}}\cap H_{b_1,\cdots,b_{l_2}}}(Sym(\zeta))^2.
\end{eqnarray}
It is known that \cite{ItoyamaNPB2018}
\begin{eqnarray}\label{aristo}
\sum_{\zeta\vdash l}Sym(\zeta)=\sum_{\Delta_1,\Delta_2,\Delta_3\vdash l}
\frac{1}{|\mathcal{T}_{\Delta_1}|^2}\mathcal{N}^H(\Delta_1,\Delta_2,\Delta_3),
\end{eqnarray}
where~$\mathcal{N}^H(\Delta_1,\Delta_2,\Delta_3)$ are the Hurwitz numbers,
and~$\Delta_i$, $i=1,2,3$, are Young diagrams of size $l$.

Similarly, we have
\begin{eqnarray}\label{hurw1}
\sum_{\zeta\in H_{b_1,\cdots,b_{l_1}}}(Sym(\zeta))^2
&=&\sum_{\zeta}|\mathcal{T}_\zeta|(Sym(\zeta))^2
=l!\sum_{\zeta}Sym(\zeta)\nonumber\\
&=&l!\sum_{\Delta_1}\sum_{\Delta_2,\Delta_3\vdash l}
\frac{1}{|\mathcal{T}_{\Delta_1}|^2}\mathcal{N}^H(\Delta_1,\Delta_2,\Delta_3),
\end{eqnarray}
\begin{eqnarray}\label{hurw2}
\sum_{\zeta\in H_{a_1,\cdots,a_{l_1}}\cap H_{b_1,\cdots,b_{l_2}}}(Sym(\zeta))^2
=l!\sum_{\Delta_1'}\sum_{\Delta_2,\Delta_3\vdash l}
\frac{1}{|\mathcal{T}_{\Delta_1'}|^2}\mathcal{N}^H(\Delta_1',\Delta_2,\Delta_3),
\end{eqnarray}
where~$\zeta$ on the right-hand side of the first line stands for every conjugacy class in $H_{b_1,\cdots,b_{l_1}}$.
The sums of~$\Delta_1$ in (\ref{hurw1}) and $\Delta_1'$ in (\ref{hurw2}) go over all Young diagrams in $H_{b_1,\cdots,b_{l_1}}$
and $H_{a_1,\cdots,a_{l_1}}\cap H_{b_1,\cdots,b_{l_2}}$, respectively.

Thus, in terms of Hurwitz numbers, (\ref{openum}) can be expressed as
\begin{eqnarray}\label{indeope}
\tilde{\sharp}_{l}&=&\sum_{l_1=2}^{min\{\sharp_L,l\}}
\sum_{\substack{b_1+\cdots+b_{l_1}=l\\b_1,\cdots,b_{l_1}\geqslant 1}}
\sum_{\Delta_1}\sum_{\Delta_2,\Delta_3\vdash l}\binom{\sharp_L}{l_1}
\frac{2\sharp_Ll!}{|H_{b_1,\cdots,b_{l_1}}||\mathcal{T}_{\Delta_1}|^2}\mathcal{N}^H(\Delta_1,\Delta_2,\Delta_3)
+\sum_{\Delta_1,\Delta_2,\Delta_3\vdash l}\frac{\sharp_L^2}{|\mathcal{T}_{\Delta_1}|^2}\mathcal{N}^H(\Delta_1,\Delta_2,\Delta_3)\nonumber\\
&&+\sum_{l_1,l_2=2}^{min\{\sharp_L,l\}}\sum_{\substack{a_1+\cdots+a_{l_1}=l\\a_1,\cdots,a_{l_1}\geqslant 1}}
\sum_{\substack{b_1+\cdots+b_{l_2}=l\\b_1,\cdots,b_{l_2}\geqslant 1}}
\sum_{\Delta_1'}\sum_{\Delta_2,\Delta_3\vdash l}
\binom{\sharp_L}{l_1}\binom{\sharp_L}{l_2}\frac{l!^2}{|H_{a_1,\cdots,a_{l_1}}||H_{b_1,\cdots,b_{l_2}}||\mathcal{T}_{\Delta_1'}|^2}
\mathcal{N}^H(\Delta_1',\Delta_2,\Delta_3).
\end{eqnarray}

When~$\sharp_L=1$, the invariants (\ref{rsigma}) are in fact the ones in the Aristotelian model \cite{ItoyamaNPB2018}.
We see that the number of the independent operators (\ref{indeope}) reduces to (\ref{aristo}).

When $\sharp_L=1$ and $\sigma_1=(1,\cdots,a_1),\sigma_2=\sigma_{3}=id$,
the invariants (\ref{rsigma}) are the ones in the red rainbow model.
In this case, the number of the independent operators (\ref{indeope})
becomes $\tilde{\sharp}_{l}=\sum_{\Delta\vdash l}\mathcal{N}^H(\Delta,[1^{a_1}],\Delta)$.

Since independent operators include disconnected and connected operators,
the relation between $\tilde{\sharp}_{l}$ and $\tilde{\sharp}_{l}^{conn}$ is
\begin{eqnarray}\label{m+n-con-ope}
\check{\eta}(q)=1+\sum_{l=1}^{\infty}\tilde{\sharp}_{l}q^{l}=PE(\check{\eta}^{conn}(q))=
PE(\sum_{l=1}^{\infty}\tilde{\sharp}_{l}^{conn}q^{l})=\prod_{l=1}^{\infty}
\frac{1}{(1-q^{l})^{\tilde{\sharp}_{l}^{conn}}}.
\end{eqnarray}
Hence the number~$\tilde{\sharp}_{l}^{conn}$ can be calculated from the plethystic logarithm
\begin{eqnarray}
PLog(\check{\eta}(q))=\sum_{l=1}^{\infty}\tilde{\sharp}_{l}^{conn}q^l=\sum_{m=1}^{\infty}
\frac{\mu(m)}{m}log\check{\eta}(q^m),
\end{eqnarray}
where~$\mu(m)$ is the M\"{o}bius function
\begin{equation}
\mu(m)=\left\{
\begin{aligned}
&0,\quad &m \emph{ has at least one repeated prime factor,}\\
&1,\quad &m=1,\\
&(-1)^n, \quad &m\emph{ is a product of n distinct primes.}\\
\end{aligned}
\right.
\end{equation}

\section{Correspondence between gauge-invariant operators and colored Dessins}

Let us rewrite the operators (\ref{rsigma}) as
\begin{eqnarray}
\mathcal{R}_{\tilde{\sigma}}^{(\vec{a},\vec{b})}
=\mathcal{M}_{I_1,I_2,I_3}\tilde{\mathcal{M}}^{I_1,\sigma_2
\circ\sigma_1^{-1}(I_2),\sigma_3\circ\sigma_1^{-1}(I_3)}
:=\mathcal{\tilde{R}}_{(id,\sigma_2\circ\sigma_1^{-1},\sigma_3\circ
\sigma_1^{-1})}^{(\vec{a},\vec{b}),\sigma_1},
\end{eqnarray}
where~$\tilde{\mathcal{M}}^{I_1,\sigma_2\circ\sigma_1^{-1}(I_2),\sigma_3\circ\sigma_1^{-1}(I_3)}$ is a rearrangement of~(\ref{barmsigma}) by~$\sigma_1$.
More precisely, for $\bar{\mathcal{M}}^{\sigma_1(I_1),\sigma_2(I_2),\sigma_3(I_3)}$, we pick out the field with~$i_1^a$ as the $a$-th element in
~$\tilde{\mathcal{M}}^{I_1,\sigma_2\circ\sigma_1^{-1}(I_2),\sigma_3\circ\sigma_1^{-1}(I_3)}$.

For colored Dessins, they are Grothendieck's Dessins d'Enfants equipped with the additional data of a coloring of the edges \cite{Galois}.
Let us establish a one-to-one correspondence between connected operators and colored Dessins with $\sharp_L^2$ colors as follows.
$\sigma_2\circ\sigma_1^{-1}$ in $\tilde{\mathcal{M}}^{I_1,\sigma_2\circ\sigma_1^{-1}(I_2),\sigma_3\circ\sigma_1^{-1}(I_3)}$
can be assigned to a Dessin by labelling the edges and going around the black vertices.
$ \sigma_3\circ\sigma_1^{-1}$ describes the permutation of the edges around the white vertices.
If the~$j$-th field in~$\mathcal{M}_{I_1,I_2,I_3}$ is~$T^l$, and the $ \sigma_3\circ\sigma_1^{-1}(j)$-th field
in~$\tilde{\mathcal{M}}^{I_1,\sigma_2\circ\sigma_1^{-1}(I_2),\sigma_3\circ\sigma_1^{-1}(I_3)}$ is~$\bar{T}_{l'}$,
then for the~$j$-th edge coming out of the white vertex, we draw it color-$(l'+(l-1)\sharp_L)$.

\begin{figure}[H]
\centering
\includegraphics[height=6.5cm]{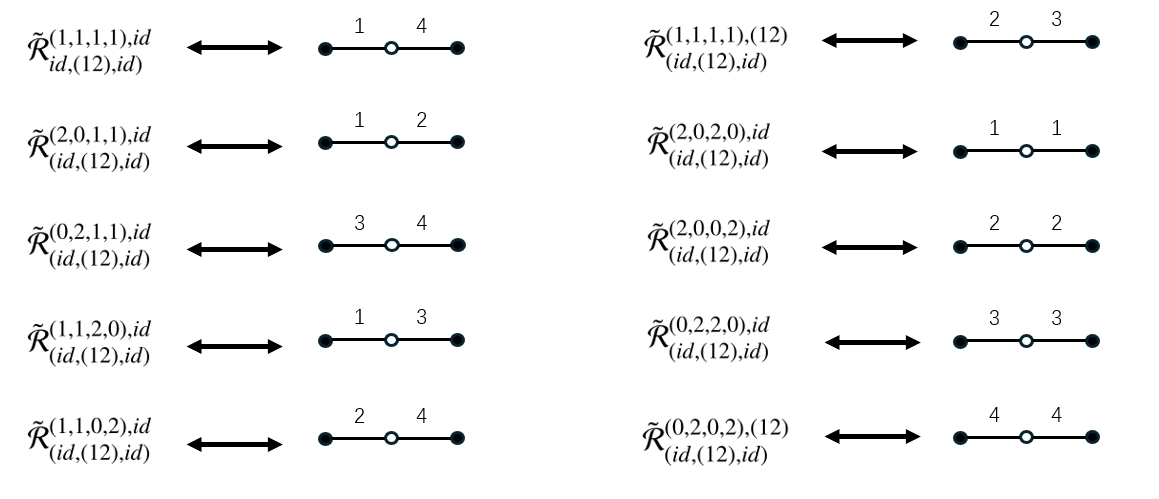}
\caption{Correspondence between some connected operators and colored Dessins.
Number above edge represents the color.}
\label{dessinconn1}
\end{figure}

For example, we list the correspondence between some connected operators with level~$2$ and colored Dessins in Fig.\ref{dessinconn1}.

The number of edges in a colored Dessin is equal to the level of the corresponding operator.
Since the number of connected operators $\tilde{\sharp}_{l}^{conn}$ with level~$l$ is given by (\ref{m+n-con-ope}),
the number of colored Dessins with $l$ edges and $\sharp_L^2$ colors is equal to~$\tilde{\sharp}_{l}^{conn}$.
For the disconnected operator consisting of~$i$ connected operators, it corresponds to~$i$-colored Dessins.

\section {Two rainbow tensor models}
\subsection {A rainbow tensor model}
By means of the connected operators $\mathcal{R}_{\tilde{\sigma}}^{(\vec{a},\vec{b})}$,
we construct a rainbow tensor model with rank-$3$
\begin{eqnarray}\label{rainbowr3}
\mathcal{Z}_{I}&=&\frac{1}{\tilde{\mathcal{Z}}_{I}}
\int \prod_{j=1}^{\sharp_L}dT^j d\bar{T}_j\exp(-\sum_{j=1}^{\sharp_L}\Tr T^j\bar{T}_j
+\sum_{a=1}^{\infty}\sum_{level (\mathcal{R}_{\tilde{\sigma}}^{(\vec{a},\vec{b})})=a}
t_{\tilde{\sigma}}^{(\vec{a},\vec{b})}\mathcal{R}_{\tilde{\sigma}}^{(\vec{a},\vec{b})}),
\end{eqnarray}
where
\begin{eqnarray}
\tilde{\mathcal{Z}}_{I}&=&\int \prod_{j=1}^{\sharp_L}dT^j d\bar{T}_j\exp(-\sum_{j=1}^{\sharp_L}\Tr T^j\bar{T}_j).
\end{eqnarray}
By requiring that the partition function (\ref{rainbowr3}) is invariant under the transformation $T^j\rightarrow T^j+\delta T^j$,
where $\delta T^j=\dsum_{c=1}^{\infty}\dsum_{level (\mathcal{R}_{\tilde{\alpha}})=c}t_{\tilde{\alpha}}^{(\vec{c},\vec{d})}\dfrac{\partial
\mathcal{R}_{\tilde{\alpha}}^{(\vec{c},\vec{d})}}{\partial \bar T_j}$,
$(\vec{c},\vec{d})=(c_1,\cdots,c_{\sharp_L},d_1,\cdots,d_{\sharp_L})$, we have

\begin{eqnarray}\label{ward3}
&&\int \prod_{j=1}^{\sharp_L}dT^j d\bar{T}_j\Big[-\dsum_{c=1}^{\infty}\dsum_{level (\mathcal{R}_{\tilde{\alpha}}^{(\vec{c},\vec{d})})=c}
level (\mathcal{R}_{\tilde{\alpha}}^{(\vec{c},\vec{d})})t_{\tilde{\alpha}}^{(\vec{c},\vec{d}) }\mathcal{R}_{\tilde{\alpha}}^{(\vec{c},\vec{d})}
+\sum_{c=1}^{\infty}\dsum_{level (\mathcal{R}_{\tilde{\alpha}}^{(\vec{c},\vec{d})})=c}\sum_{j=1}^{\sharp_L}
t_{\tilde{\alpha}}^{(\vec{c},\vec{d})}\Delta_{I,j} \mathcal{R}_{\tilde{\alpha}}^{(\vec{c},\vec{d})}\nonumber\\
&&+\sum_{j=1}^{\sharp_L}\dsum_{a,c=1}^{\infty}\dsum_{level (\mathcal{R}_{\tilde{\sigma}}^{(\vec{a},\vec{b})})=a}
\dsum_{level (\mathcal{R}_{\tilde{\alpha}}^{(\vec{c},\vec{d}) })=c}t_{\tilde{\sigma}}^{(\vec{a},\vec{b})}t_{\tilde{\alpha}}^{(\vec{c},\vec{d})}
\{\mathcal{R}_{\tilde{\sigma}}^{(\vec{a},\vec{b})},\mathcal{R}_{\tilde{\alpha}}^{(\vec{c},\vec{d})}\}_{I,j}\Big]
\cdot \exp(\sum_{a=1}^{\infty}\sum_{level (\mathcal{R}_{\tilde{\sigma}}^{(\vec{a},\vec{b})})=a}
t_{\tilde{\sigma}}^{(\vec{a},\vec{b})}\mathcal{R}_{\tilde{\sigma}}^{(\vec{a},\vec{b})}\nonumber\\
&&-\sum_{j=1}^{\sharp_L}\Tr T^j\bar{T}_j)=0,
\end{eqnarray}
where $\{,\}_{I,j}$ and~$\Delta_{I,j}$ are respectively the cut and join operations on the gauge-invariant operator
$\mathcal{R}_{\tilde{\sigma}}^{(\vec{a},\vec{b})}$.
The actions of the cut operations on the gauge-invariant operator $\mathcal{R}_{\tilde{\alpha}}^{(\vec{a},\vec{b})}$ are
\begin{eqnarray}\label{cut}
&&\Delta_{I,j} \mathcal{R}_{\tilde{\alpha}}^{(\vec{a},\vec{b})}
=\sum_{i_1=1}^{N_{1}}
\sum_{i_2=1}^{N_{2}}\sum_{i_{3}=1}^{N_{3}}
\dfrac{\partial^2 \mathcal{R}_{\tilde{\alpha}}^{(\vec{a},\vec{b})}}{\partial T^j_{i_1,i_2,i_{3}}
\partial \bar{T}_j^{i_1,i_2,i_{3}}}\nonumber\\
&&
=\sum_{k=1}^3\sum_{\Gamma_1}
\sum_{\tilde{\beta}_1,\cdots,\tilde{\beta}_k}
(\Delta_j)_{(\tilde{\alpha},\tilde{\beta}_1,\cdots,\tilde{\beta}_k)}
^{\bar\mu}
\mathcal{R}_{\tilde{\beta}_1}^{(\vec{a}^{(1)},\vec{b}^{(1)})}\cdots
\mathcal{R}_{\tilde{\beta}_k}^{(\vec{a}^{(k)},\vec{b}^{(k)})},
\end{eqnarray}
where $\Gamma_1= \{\vec{a}^{(1)}+\cdots+\vec{a}^{(k)}+\vec{(1)}_j=\vec{a},
 \vec{b}^{(1)}+\cdots+\vec{b}^{(k)}+\vec{(1)}_j=\vec{b},
 a_1^{(1)}\leq \cdots\leq a_1^{(k)}\}$, $\sum_{i=1}^{\sharp_L}a_i\geqslant 2$,
 $\vec{(1)}_j:=(\overbrace{ 0,\cdots,0}^{j-1},1,\overbrace{ 0,\cdots,0}^{\sharp_L-j})$,
 $\bar\mu=((\vec{a},\vec{b}),(\vec{a}^{(1)},\vec{b}^{(1)}),\cdots,(\vec{a}^{(k)},\vec{b}^{(k)}))$,
 and
$(\Delta_j)_{(\tilde{\alpha},\tilde{\beta}_1,\cdots,\tilde{\beta}_k)}
^{\bar\mu}$
are polynomials of $N_i$.

The actions of the join operations on the gauge invariant operators
$\mathcal{R}_{\tilde{\alpha}}^{(\vec{a},\vec{b})}$ and $\mathcal{R}_{\tilde{\beta}}^{(\vec{c},\vec{d})}$ are
\begin{eqnarray}\label{join}
\{\mathcal{R}_{\tilde{\alpha}}^{(\vec{a},\vec{b})},\mathcal{R}_{\tilde{\beta}}^{(\vec{c},\vec{d})}\}_{I,j}
&=&
\sum_{i_1=1}^{N_{1}}
\sum_{i_2=1}^{N_{2}}\sum_{i_{3}=1}^{N_{3}}
\dfrac{\partial \mathcal{R}_{\tilde{\alpha}}^{(\vec{a},\vec{b})}}{\partial T^j_{i_1,i_2,i_{3}}}\dfrac{\partial
\mathcal{R}_{\tilde{\beta}}^{(\vec{c},\vec{d})}}{\partial \bar{T}_j^{i_1,i_2,i_{3}}}\nonumber\\
&=&
\sum_{\Upsilon_1}
({\Lambda}_j)_{(\tilde{\alpha},\tilde{\beta},\tilde{\gamma})}
^{((\vec{a},\vec{b}),(\vec{c},\vec{d}),\bar\xi)}
\mathcal{R}_{\tilde{\gamma}}^{\bar\xi},
\end{eqnarray}
where~$\Upsilon_1=\{\tilde{\gamma},level(\mathcal{R}_{\tilde{\gamma}}^{\bar\xi})=level(\mathcal{R}_{\tilde{\alpha}}^{(\vec{a},\vec{b})})
+level(\mathcal{R}_{\tilde{\beta}}^{(\vec{c},\vec{d})})-1\}$,~$\bar\xi=(\vec{a}+\vec{c}-\vec{(1)}_j,\vec{b}+\vec{d}-\vec{(1)}_j)$ and
$({\Lambda}_j)_{(\tilde{\alpha},\tilde{\beta},\tilde{\gamma})}^{((\vec{a},\vec{b}),(\vec{c},\vec{d}), \bar\xi)}$
are integer coefficients.

It should be noted that the connected operators $\mathcal{R}_{\tilde{\sigma}}^{(\vec{a},\vec{b})}$ in (\ref{rainbowr3})
are in the ring of invariants generated by
$\mathcal{R}_{(id,(12),id)}^{(2,\overbrace{\scriptstyle 0,\cdots,0}^{\sharp_L-1},2,\overbrace{\scriptstyle 0,\cdots,0}^{\sharp_L-1})}$,
$\mathcal{R}_{((12),id,id)}^{(2,\overbrace{\scriptstyle 0,\cdots,0}^{\sharp_L-1},2,\overbrace{\scriptstyle 0,\cdots,0}^{\sharp_L-1})}$,
$\mathcal{R}_{(id,id,id)}^{(1,\overbrace{\scriptstyle 0,\cdots,0}^{\sharp_L-1+i},1,\overbrace{\scriptstyle 0,\cdots,0}^{\sharp_L-i})}$
and $\mathcal{R}_{(id,id,id)}^{(\overbrace{\scriptstyle 0,\cdots,0}^{i},1,\overbrace{\scriptstyle 0,\cdots,0}^{\sharp_L-1+i},
1,\overbrace{\scriptstyle 0,\cdots,0}^{\sharp_L-1})},i=1,\cdots,\sharp_L-1$,
with addition, multiplication, cut and join operations.

From (\ref{ward3}), it is not difficult to obtain that
\begin{eqnarray}\label{dw}
 \hat{D}\mathcal{Z}_{I}=\mathcal{\hat{W}}_{I}\mathcal{Z}_{I},
\end{eqnarray}
where
\begin{eqnarray}\label{operatorD}
\hat{D}=\sum_{c=1}^{\infty}\dsum_{level (\mathcal{R}_{\tilde{\alpha}}^{(\vec{c},\vec{d})})=c}level (\mathcal{R}_{\tilde{\alpha}}^{(\vec{c},\vec{d})})
t_{\tilde{\alpha}}^{(\vec{c},\vec{d}) }\frac{\partial}{\partial t_{\tilde{\alpha}}^{(\vec{c},\vec{d})}},
\end{eqnarray}
\begin{eqnarray}\label{operator3}
\mathcal{\hat{W}}_{I}
&=&\sum_{c=1}^{\infty}\dsum_{level (\mathcal{R}_{\tilde{\alpha}}^{(\vec{c},\vec{d})})=c}
\sum_{k=1}^3\sum_{j=1}^{\sharp_L}\sum_{\Gamma_2}
\sum_{\tilde{\beta}_1,\cdots,\tilde{\beta}_k}(1-\delta_{c_j,1})
(\Delta_j)_{(\tilde{\alpha},\tilde{\beta}_1,\cdots,\tilde{\beta}_k)}
^{\bar\nu}t_{\tilde{\alpha}}^{(\vec{c},\vec{d})}
\frac{\partial}{\partial t_{\tilde{\beta}_1}^{(\vec{c}^{(1)},\vec{d}^{(1)})}}\cdots
\frac{\partial}{\partial t_{\tilde{\beta}_k}^{(\vec{c}^{(k)},\vec{d}^{(k)})}}
\nonumber\\
&+&\sum_{a,c=1}^{\infty}\sum_{level (\mathcal{R}_{\tilde{\sigma}}^{(\vec{a},\vec{b})})=a}
\dsum_{level (\mathcal{R}_{\tilde{\alpha}}^{(\vec{c},\vec{d})})=c}\dsum_{\Upsilon_2}
\sum_{j=1}^{\sharp_L}{(\Lambda_j)}_{(\tilde{\sigma},\tilde{\alpha},\tilde{\beta})}
^{((\vec{a},\vec{b}),(\vec{c},\vec{d}),\bar\xi)}
t_{\tilde{\sigma}}^{(\vec{a},\vec{b})}t_{\tilde{\alpha}}^{(\vec{c},\vec{d})}
\frac{\partial}{\partial t_{\tilde{\beta}}^{\bar\xi}}\nonumber\\
&+&\sum_{j=1}^{\sharp_L}t_{(id,id,id)}^{(\overbrace{\scriptstyle 0,\cdots,0}^{j-1},1,
\overbrace{\scriptstyle 0,\cdots,0}^{\sharp_L-1},1,\overbrace{\scriptstyle 0,\cdots,0}^{\sharp_L-j})}
N_{1}N_{2}N_{3},
\end{eqnarray}
in which $\Gamma_2= \{ \sum_{i=1}^kc^{(i)}_j+1=c_j, \sum_{i=1}^kd^{(i)}_j+1=d_j,
\sum_{i=1}^kc^{(i)}_m=c_m,\sum_{i=1}^kd^{(i)}_n=d_n,m,n\neq j,c_1^{(1)}\leq \cdots\leq c_1^{(k)}\}$,
$\Upsilon_2=\{\tilde{\beta},level(\mathcal{R}_{\tilde{\beta}}^{\bar\xi})=level (\mathcal{R}_{\tilde{\sigma}}^{(\vec{a},\vec{b})})
+level (\mathcal{R}_{\tilde{\alpha}}^{(\vec{c},\vec{d})})-1\}$,
$\bar\nu=((\vec{c},\vec{d}),(\vec{c}^{(1)},\vec{d}^{(1)}),\cdots,(\vec{c}^{(k)},\vec{d}^{(k)}))$,
and $\tilde{\alpha}$, $\tilde{\beta}$,
$\tilde{\beta}_k$ and $\tilde{\sigma}$ are taken from indices of connected operators of rainbow model with rank-$3$.

The operators~$\hat{D}$ and~$\mathcal{\hat{W}}_{I}$ satisfy
\begin{eqnarray}\label{}
[\hat{D},\mathcal{\hat{W}}_{I}]=\mathcal{\hat{W}}_{I}.
\end{eqnarray}
Let us rewrite (\ref{rainbowr3}) as
\begin{eqnarray}
\mathcal{Z}_{I}=\sum_{s=0}^{\infty}\mathcal{Z}_{I}^{(s)},
\end{eqnarray}
where
\begin{eqnarray}
\mathcal{Z}_{I}^{(s)}&=&\sum_{l=0}^{\infty}\sum_{a_1+\cdots +a_l=s}
\sum_{\substack{level (\mathcal{R}_{\tilde{\sigma}_i}^{(\vec{a}^{(i)},\vec{b}^{(i)})})=a_i,\\i=1,\cdots,l}}
\frac{1}{l!}\langle\langle \prod_{i=1}^{l}\mathcal{R}_{\tilde{\sigma}_i}^{(\vec{a}^{(i)},\vec{b}^{(i)})}\rangle\rangle_{I,\sharp_L}
\prod_{i=1}^{l}t_{\tilde{\sigma}_i}^{(\vec{a}^{(i)},\vec{b}^{(i)})},
\end{eqnarray}
in which
\begin{eqnarray}\label{}
\langle \langle \prod_{i=1}^{l}\mathcal{R}_{\tilde{\sigma}_i}^{(\vec{a}^{(i)},\vec{b}^{(i)})}\rangle\rangle_{I,\sharp_L}
&\equiv&\frac{1}{\tilde{\mathcal{Z}}_{I}}\int \dprod_{j=1}^{\sharp_L}dT^j d\bar{T}_j\mathcal{R}_{\tilde{\sigma}_1}^{(\vec{a}^{(1)},\vec{b}^{(1)})}
 \cdots\mathcal{R}_{\tilde{\sigma}_l}^{(\vec{a}^{(l)},\vec{b}^{(l)})}\exp(- \sum_{j=1}^{\sharp_L}\Tr T^j\bar{T}_j).
\end{eqnarray}

Then the operators $\hat D$ and $\mathcal{\hat{W}}_{I}$ acting on $\mathcal{Z}_{I}^{(s)}$ give
\begin{eqnarray}\label{}
&&\hat {D} \mathcal{Z}_{I}^{(s)}=s\mathcal{Z}_{I}^{(s)},\nonumber\\
&&\mathcal{\hat{W}}_{I}\mathcal{Z}_{I}^{(s)}=(s+1)\mathcal{Z}_{I}^{(s+1)}.
\end{eqnarray}
We see that the operators~$\hat {D}$ and $\mathcal{\hat{W}}_{I}$ are indeed the operators preserving and increasing the grading, respectively.
Thus the partition function (\ref{rainbowr3}) can be realized by the $W$-representation
\begin{eqnarray}\label{exp3}
\mathcal{Z}_{I}=\exp(\mathcal{\hat{W}}_{I})\cdot 1.
\end{eqnarray}

Let us write the $k$-th power of the operator $\mathcal{\hat{W}}_{I}$ as
\begin{eqnarray}\label{wm}
&&\mathcal{\hat{W}}_{I}^k
=\sum_{i=1}^{2k}\sum_{j=1}^{3k}
\sum_{\Upsilon_3}
{P}_{\tilde{\alpha}_1,\cdots,\tilde{\alpha}_i;(\vec{a}^{(1)},\vec{b}^{(1)}),\cdots,(\vec{a}^{(i)},\vec{b}^{(i)})}
^{ \tilde{\beta}_1,\cdots,\tilde{\beta}_j;(\vec{c}^{(1)},\vec{d}^{(1)}),\cdots,(\vec{c}^{(j)},\vec{d}^{(j)})}
t_{\tilde{\alpha}_1}^{(\vec{a}^{(1)},\vec{b}^{(1)})}\cdots t_{\tilde{\alpha}_{i}}^{(\vec{a}^{(i)},\vec{b}^{(i)})}
\nonumber\\
&&\cdot\frac{\partial}{\partial {t_{\beta_1}^{(\vec{c}^{(1)},\vec{d}^{(1)})}}}
\cdots \frac{\partial}{\partial t_{\beta_{j}}^{(\vec{c}^{(j)},\vec{d}^{(j)})}}
+\sum_{i=1}^{k}\sum_{\Upsilon_4}
{P}_{~\tilde{\alpha}_1,\cdots,\tilde{\alpha}_i}
^{(\vec{a}^{(1)},\vec{b}^{(1)}),\cdots,(\vec{a}^{(i)},\vec{b}^{(i)})}
t_{\tilde{\alpha}_1}^{(\vec{a}^{(1)},\vec{b}^{(1)})}
\cdots  t_{\tilde{\alpha}_{i}}^{(\vec{a}^{(i)},\vec{b}^{(i)})},\nonumber\\
\end{eqnarray}
where $\Upsilon_3=\{level(\mathcal{R}_{\tilde{\alpha}_1}^{(\vec{a}^{(1)},\vec{b}^{(1)})})
+\cdots+level(\mathcal{R}_{\tilde{\alpha}_i}^{(\vec{a}^{(i)},\vec{b}^{(i)})})
= level(\mathcal{R}_{\tilde{\beta}_1}^{(\vec{c}^{(1)},\vec{d}^{(1)})})+\cdots
+level(\mathcal{R}_{\tilde{\beta}_j}^{(\vec{c}^{(j)},\vec{d}^{(j)})})+k\}$,
$\Upsilon_4=\{level(\mathcal{R}_{\tilde{\alpha}_1}^{(\vec{a}^{(1)},\vec{b}^{(1)})})+\cdots
+level(\mathcal{R}_{\tilde{\alpha}_i}^{(\vec{a}^{(i)},\vec{b}^{(i)})})=k\}$,
 ${P}_{\tilde{\alpha}_1,\cdots,\tilde{\alpha}_i;(\vec{a}^{(1)},\vec{b}^{(1)}),\cdots,(\vec{a}^{(i)},\vec{b}^{(i)})}
^{ \tilde{\beta}_1,\cdots,\tilde{\beta}_j;(\vec{c}^{(1)},\vec{d}^{(1)}),\cdots,(\vec{c}^{(j)},\vec{d}^{(j)})}$  and ${P}_{~\tilde{\alpha}_1,\cdots,\tilde{\alpha}_i}^{(\vec{a}^{(1)},\vec{b}^{(1)}),\cdots,(\vec{a}^{(i)},\vec{b}^{(i)})}$
are polynomials of $N_{1}$, $N_{2}$ and $N_{3}$.

Then we may derive the compact expression of correlators from (\ref{exp3})
\begin{eqnarray}\label{corrf3}
\langle \langle \prod_{i=1}^{l}\mathcal{R}_{\tilde{\sigma}_i}^{(\vec{a}^{(i)},\vec{b}^{(i)})}\rangle\rangle_{I,\sharp_L}
&=&\frac{l!}{k!\lambda_{(\tilde{\sigma}_1,\cdots,\tilde{\sigma}_l)}}
\sum_{\tau}{P}_{~\tau(\tilde{\sigma}_1),\cdots,\tau(\tilde{\sigma}_l)}
^{(\vec{a}^{(1)},\vec{b}^{(1)}),\cdots,
(\vec{a}^{(l)},\vec{b}^{(l)})},
\end{eqnarray}
where~$k=\sum_{i=1}^{\sharp_L}\sum_{j=1}^{l}a_i^{(j)}$,
$\tau$ denotes all distinct permutations of $(\tilde{\sigma}_1,\cdots,\tilde{\sigma}_l)$
and~$\lambda_{(\tilde{\sigma}_1,\cdots,\tilde{\sigma}_l)}$
is the number of $\tau$ with respect to~$\tilde{\sigma}_1,\cdots,\tilde{\sigma}_l$.
It is obvious that correlators are zero unless
$\sum_{j=1}^{l}a_i^{(j)}=\sum_{j=1}^{l}b_i^{(j)},i=1,\cdots,\sharp_L$.

We have established a one-to-one correspondence between connected operators and colored Dessins.
With the help of Dessins, we may derive another compact expression of correlators
\begin{eqnarray}\label{wickth}
&&\langle\langle\mathcal{R}_{\tilde{\sigma}}^{(a_1,\cdots,a_{\sharp_L},b_1,\cdots,b_{\sharp_L})}\rangle\rangle_{I,\sharp_L}
=\langle\langle\prod_{i=1}^{l}\mathcal{R}_{\tilde{\sigma}_i}
^{(a_1^{(i)},\cdots,a_{\sharp_L}^{(i)},b_1^{(i)},\cdots,b_{\sharp_L}^{(i)})}\rangle\rangle_{I,\sharp_L}
=\langle\langle\mathcal{\tilde{R}}_{(id,\sigma_2\circ\sigma_1^{-1},\sigma_3\circ\sigma_1^{-1})}
^{(a_1,\cdots,a_{\sharp_L},b_1,\cdots,b_{\sharp_L}),\sigma_1}\rangle\rangle_{I,\sharp_L}\nonumber\\
&=&\sum_{\gamma\in S_{a_1}\times \cdots\times S_{a_{\sharp_L}}}
N_1^{\sharp(\gamma\circ\sigma_1)}N_2^{\sharp(\gamma\circ\sigma_2)}N_3^{\sharp(\gamma\circ\sigma_3)}\nonumber\\
&=&\sum_{\gamma\in S_{a_1}\times \cdots\times S_{a_{\sharp_L}}}\sum_{\tau_1,\tau_2,\tau_3\in S_{a_1+ \cdots+a_{\sharp_L}}}
\delta(\sigma_1\circ\gamma\circ\tau_1)\delta(\sigma_2\circ\gamma\circ\tau_2)\delta(\sigma_3\circ\gamma\circ\tau_3)
N_1^{\sharp(\tau_1)}N_2^{\sharp(\tau_2)}N_3^{\sharp(\tau_3)}\nonumber\\
&=&\frac{1}{\mathcal{N}_1}\sum_{\Theta_1}\sum_{\Theta_2}
\sum_{\tau_1,\tau_2,\tau_3\in S_{a_1+\cdots+a_{\sharp_L}}} \frac{1}{(a_1!\cdots a_{\sharp_L}!)^3}
\delta(\gamma\circ\tau_1)\delta(\sigma_2\circ\sigma_1^{-1}\circ\gamma\circ\tau_2)
\delta(\sigma_3\circ\sigma_1^{-1}\circ\gamma\circ\tau_3)\nonumber\\
&&
\cdot
 N_1^{\sharp(\gamma)}
N_2^{\sharp(\tau_2)+\sharp{(\sigma_2}\circ\sigma_1^{-1})-\sum_{i=1}^{\sharp_L}a_i}
N_3^{\sharp(\tau_3)+\sharp{(\sigma_3}\circ\sigma_1^{-1})-\sum_{i=1}^{\sharp_L}a_i}\nonumber\\
&=&\frac{1}{\mathcal{N}_1}
 \sum_{D_1,D_2,D_3\in{D_{col}}}\frac{N_1^{\frac{1}{2}\chi(D_1)}
 N_2^{\chi(D_2)-\frac{1}{2}\chi(D_1)}N_3^{\chi(D_3)-\frac{1}{2}\chi(D_1)}}{|Aut(D_{1})|\cdot |Aut(D_{2})|\cdot |Aut(D_{3})|},
\end{eqnarray}
where we  have used the Wick theorem in the second line, $\sharp(\alpha)$ is the number of cycles in the permutation $\alpha$,
and $[\alpha]_{a_1,\cdots,a_{\sharp_L}}$ is the set of permutations in $S_{a_1+\cdots+a_{\sharp_L}}$
related to $\alpha$ by conjugation with elements in~$S_{a_1}\times \cdots\times S_{a_{\sharp_L}}$,
$\mathcal{N}_1=\frac{1}{(a_1!\cdots a_{\sharp_L}!)^3} \mid[\sigma_2]_{a_1,\cdots,a_{\sharp_L}}\mid
\cdot \mid [\sigma_3]_{a_1,\cdots,a_{\sharp_L}}\mid
N_2^{\sharp{(\sigma_2\circ\sigma_1^{-1})}-\sum_{i=1}^{\sharp_L}a_i}
N_3^{\sharp{(\sigma_3\circ\sigma_1^{-1})}-\sum_{i=1}^{\sharp_L}a_i}$,
 $\Theta_1=\{ \sigma_2\in [\sigma_2]_{a_1,\cdots,a_{\sharp_L}},\sigma_3\in [\sigma_3]_{a_1,\cdots,a_{\sharp_L}}\}$ and
 $\Theta_2=\{\gamma,\sigma_1^{-1}\circ\gamma\in S_{a_1}\times \cdots\times S_{a_{\sharp_L}}\}$.

The formula (\ref{wickth}) shows a way to calculate correlators graphically. For the choice of a contraction~$\gamma$, we choose a coloring of
the cyclically ordered edges coming out of a white vertex of three Dessins~$D_1,D_2,D_3$. Namely,~$\gamma$ is associated with~$\sigma_1$ in Belyi
literature conventions. By attaching black vertices to any edge coming out of white vertices, we obtain the colored Dessin~$D_1$ in the last line
of (\ref{wickth}). There are~$\sharp_L$ colors in colored Dessin~$D_1$, which represent~$\sharp_L$ types of contractions between $\sharp_L$ complex tensors
$T_{i_{1}, \cdots, i_{r}}^{j}$ and their complex conjugates, respectively. $\sigma_2\circ\sigma_1^{-1}$ is assigned to the Dessin~$D_2$ by labelling
the edges and going around the black vertices. There are also~$\sharp_L$ colors in colored Dessin~$D_2$ representing~$\sharp_L$ types of contractions.
Similarly, $\sigma_3\circ\sigma_1^{-1}$ is assigned to the Dessin~$D_3$ by labelling the edges and going around the black vertices.
Likewise, the Dessin~$D_3$ is a colored Dessin with~$\sharp_L$ colors. $Aut(D_{1})$, $Aut(D_{2})$ and $Aut(D_{3})$ are the automorphism groups of corresponding
colored Dessins \cite{{Galois}}. $\chi(D_1)$, $\chi(D_2)$ and $\chi(D_3)$ are Euler characteristics of  corresponding colored Dessins.

\subsection {A rainbow tensor model with two tensors}

Let us consider the rainbow tensor model (\ref{rainbowr3}) with two tensors~$T^1$ and~$T^2$
\begin{eqnarray}\label{rainbow23}
\mathcal{Z}_{I,2T}&=&\frac{1}{\tilde{\mathcal{Z}}_{I,2T}}
\int \prod_{j=1}^{2}dT^j d\bar{T}_j\exp(-\sum_{j=1}^{2}\Tr T^j\bar{T}_j
+\sum_{a_1+a_{2}=1}^{\infty}\sum_{level (\mathcal{R}_{\tilde{\sigma}}^{(a_1,a_{2},b_1,b_{2})})=a_1+a_{2}}
t_{\tilde{\sigma}}^{(a_1,a_{2},b_1,b_{2})}\mathcal{R}_{\tilde{\sigma}}^{(a_1,a_{2},b_1,b_{2})})
\nonumber\\
&=&\exp(\mathcal{\hat{W}}'_{I})\cdot 1,
\end{eqnarray}
where
\begin{eqnarray}
\tilde{\mathcal{Z}}_{I,2T}&=&\int \prod_{j=1}^{2}dT^j d\bar{T}_j\exp(-\sum_{j=1}^{2}\Tr T^j\bar{T}_j).
\end{eqnarray}
and
\begin{eqnarray}\label{operator23}
\mathcal{\hat{W}}'_{I}
&=&\sum_{c_1+c_{2}=1}^{\infty}\dsum_{level (\mathcal{R}_{\tilde{\alpha}}^{(c_1,c_{2},d_1,d_{2})})=c_1+c_2}
\sum_{k=1}^3\sum_{\Gamma_3}
\sum_{\tilde{\beta}_1,\cdots,\tilde{\beta}_k}(1-\delta_{c_1,1})
(\Delta_1)_{(\tilde{\alpha},\tilde{\beta}_1,\cdots,\tilde{\beta}_k)}
^{(\bar\eta,\bar\eta_1,\cdots,\bar\eta_k)}t_{\tilde{\alpha}}^{\bar\eta}
\frac{\partial}{\partial t_{\tilde{\beta}_1}^{\bar\eta_1}}\cdots
\frac{\partial}{\partial t_{\tilde{\beta}_k}^{\bar\eta_k}}
\nonumber\\
&&
\nonumber\\
&+&\sum_{c_1+c_{2}=1}^{\infty}\dsum_{level (\mathcal{R}_{\tilde{\alpha}}^{(c_1,c_{2},d_1,d_{2})})=c_1+c_2}
\sum_{k=1}^3\sum_{\Gamma_4}
\sum_{\tilde{\beta}_1,\cdots,\tilde{\beta}_k}(1-\delta_{c_2,1})
(\Delta_2)_{(\tilde{\alpha},\tilde{\beta}_1,\cdots,\tilde{\beta}_k)}
^{(\bar\eta,\bar\eta_1,\cdots,\bar\eta_k)}t_{\tilde{\alpha}}^{\bar\eta}
\frac{\partial}{\partial t_{\tilde{\beta}_1}^{\bar\eta_1}}\cdots
\frac{\partial}{\partial t_{\tilde{\beta}_k}^{\bar\eta_k}}
\nonumber\\
&&
\nonumber\\
&+&\sum_{a_1+a_{2}=1}^{\infty}\sum_{c_1+c_{2}=1}^{\infty}\dsum_{\Upsilon_5}
\dsum_{\Upsilon_6 }
{(\Lambda_1)}_{(\tilde{\sigma},\tilde{\alpha},\tilde{\beta})}^{(\bar\rho,\bar\eta,
\bar{\xi}_1)}
t_{\tilde{\sigma}}^{\bar\rho}t_{\tilde{\alpha}}^{\bar\eta}
\frac{\partial}{\partial t_{\tilde{\beta}}^{\bar{\xi}_1}}
+\sum_{a_1+a_{2}=1}^{\infty}\sum_{c_1+c_{2}=1}^{\infty}\dsum_{\Upsilon_5}
\dsum_{\Upsilon_7}
{(\Lambda_2)}_{(\tilde{\sigma},\tilde{\alpha},\tilde{\beta})}^{(\bar\rho,\bar\eta,
\bar{\xi}_2)}
t_{\tilde{\sigma}}^{\bar\rho}t_{\tilde{\alpha}}^{\bar\eta}
\frac{\partial}{\partial t_{\tilde{\beta}}^{\bar{\xi}_2}}
\nonumber\\
&+&t_{(id,id,id)}^{(1,0,1,0)}N_1N_2N_3+t_{(id,id,id)}^{(0,1,0,1)}N_1N_2N_3,
\end{eqnarray}
in which $\Gamma_3=\{ \sum_{i=1}^kc^{(i)}_1+1=c_1, \sum_{i=1}^kd^{(i)}_1+1=
d_1, \sum_{i=1}^kc^{(i)}_2=c_2,\sum_{i=1}^kd^{(i)}_2=d_2,c_1^{(1)}\leq \cdots\leq c_1^{(k)}\}$,
$\Gamma_4= \{\sum_{i=1}^kc^{(i)}_2+1=c_2,\sum_{i=1}^kd^{(i)}_2+1=d_2,
\sum_{i=1}^kc^{(i)}_1=c_1,\sum_{i=1}^kd^{(i)}_1=d_1, c_1^{(1)}\leq \cdots\leq c_1^{(k)}\}$,
$\Upsilon_5=\{level (\mathcal{R}_{\tilde{\sigma}}^{(a_1,a_{2},b_1,b_{2})})=a_1+a_2,
 level (\mathcal{R}_{\tilde{\alpha}}^{(c_1,c_{2},d_1,d_{2})})=c_1+c_2\}$,
$\Upsilon_6= \{\tilde{\beta},level(\mathcal{R}_{\tilde{\beta}}^{\bar{\xi}_1})=
level (\mathcal{R}_{\tilde{\sigma}}^{(a_1,a_{2},b_1,b_{2})})+ level (\mathcal{R}_{\tilde{\alpha}}^{(c_1,c_{2},d_1,d_{2})})-1\}$,
$\Upsilon_7= \{\tilde{\beta},level(\mathcal{R}_{\tilde{\beta}}^{\bar{\xi}_2})=
level (\mathcal{R}_{\tilde{\sigma}}^{(a_1,a_{2},b_1,b_{2})})
+ level (\mathcal{R}_{\tilde{\alpha}}^{(c_1,c_{2},d_1,d_{2})})-1\}$,
$\bar\rho=(a_1,a_{2},b_1,b_{2})$, $\bar\xi_1=(a_1+c_1-1,a_2+c_2,b_1+d_1-1,b_{2}+d_{2})$, $\bar{\xi}_2=(a_1+c_1,a_2+c_2-1,b_1+d_1,b_{2}+d_{2}-1)$,
$\bar\eta=(c_1,c_{2},d_1,d_{2})$ and $\bar\eta_i=(c^{(i)}_1,c^{(i)}_{2},d^{(i)}_1,d^{(i)}_{2}),~i=1,\cdots, k$.

The correlators are given by
\begin{eqnarray}\label{twotensor}
\langle \langle \prod_{i=1}^{l}\mathcal{R}_{\tilde{\sigma}_i}
^{(a_1^{(i)},a_{2}^{(i)},b_1^{(i)},b_{2}^{(i)})}\rangle\rangle_{I,\sharp_L=2}
&=&\frac{l!}{k!\lambda_{(\tilde{\sigma}_1,\cdots,\tilde{\sigma}_l)}}
\sum_{\tau}{P}_{~\tau(\tilde{\sigma}_1),\cdots,\tau(\tilde{\sigma}_l)}
^{(a_1^{(1)},a_{2}^{(1)},b_1^{(1)},b_{2}^{(1)}),\cdots,
(a_1^{(l)},a_{2}^{(l)},b_1^{(l)},b_{2}^{(l)})}\nonumber\\
&=&\frac{1}{\mathcal{N}_1}{\sum_{D_1,D_2,D_3\in{D_{col}}}\frac{N_1^{\frac{1}{2}\chi(D_1)}N_2^{\chi(D_2)
-\frac{1}{2}\chi(D_1)}N_3^{\chi(D_3)
-\frac{1}{2}\chi(D_1)}}{|Aut(D_{1})|\cdot |Aut(D_{2})|\cdot |Aut(D_{3})|}}.
\end{eqnarray}

Let us list some correlators from the first line of~(\ref{twotensor})
by calculating $\mathcal{\hat{W}}'_{I}$,
\begin{eqnarray}\label{correlators}
 \langle\langle \mathcal{R}_{(id,id,id)}^{(1,1,1,1)}\rangle\rangle_{I,\sharp_L=2} &=&N_1^2N_2^2N_3^2,\nonumber\\
 \langle\langle \mathcal{R}_{(id,(12),(12))}^{(1,1,1,1)}\rangle\rangle_{I,\sharp_L=2} &=&N_1^2N_2N_3,\nonumber\\
 \langle \langle\mathcal{R}_{((12),id,(12))}^{(1,1,1,1)}\rangle\rangle_{I,\sharp_L=2} &=&N_1N_2^2N_3,\nonumber\\
 \langle \langle\mathcal{R}_{((12),(12),id)}^{(1,1,1,1)}\rangle\rangle_{I,\sharp_L=2} &=&N_1N_2N_3^2,\nonumber\\
 \langle \langle\mathcal{R}_{(id,id,(12))}^{(1,1,1,1)}\rangle\rangle_{I,\sharp_L=2} &=&N_1^2N_2^2N_3,\nonumber\\
 \langle \langle\mathcal{R}_{(id,(12),id)}^{(1,1,1,1)}\rangle\rangle_{I,\sharp_L=2} &=&N_1^2N_2N_3^2,\nonumber\\
 \langle \langle\mathcal{R}_{((12),id,id)}^{(1,1,1,1)}\rangle\rangle_{I,\sharp_L=2} &=&N_1N_2^2N_3^2,\nonumber\\
 \langle \langle\mathcal{R}_{((12),(12),(12))}^{(1,1,1,1)}\rangle\rangle_{I,\sharp_L=2} &=&N_1N_2N_3.
 \end{eqnarray}

\begin{figure}[H]
\centering
\includegraphics[height=1cm]{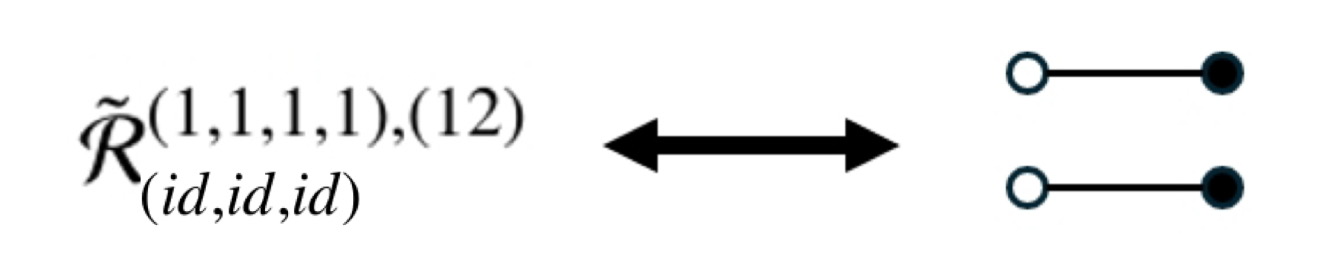}
\caption{Correspondence between $\mathcal{\tilde{R}}_{(id,id,id)}^{(1,1,1,1),(12)}$ and the Dessin.}
\label{rididid}
\end{figure}

These correlators can also be obtained by the second line of~(\ref{twotensor}).
For example, let us consider the operator
$\mathcal{R}_{((12),(12),(12))}^{(1,1,1,1)}=\mathcal{\tilde{R}}_{(id,id,id)}^{(1,1,1,1),(12)}$.
It can be assigned to the Dessin (see Fig. \ref{rididid})

\begin{figure}[H]
\centering
\includegraphics[height=1.1cm]{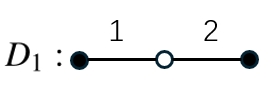}
\caption{The Dessin~$D_1$ of $\langle\langle\mathcal{\tilde{R}}_{(id,id,id)}^{(1,1,1,1),(12)}\rangle\rangle_{I,\sharp_L=2}$.}
\label{id12121}
\end{figure}

\begin{figure}[H]
\centering
\includegraphics[height=2.8cm]{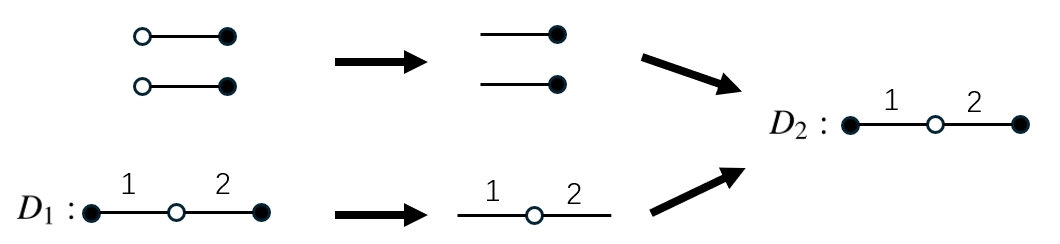}
\caption{The Dessin~$D_2$ of $\langle\langle\mathcal{\tilde{R}}_{(id,id,id)}^{(1,1,1,1),(12)}\rangle\rangle_{I,\sharp_L=2}$.}
\label{id12122}
\end{figure}

Since the only choice of the contraction~$\gamma$ is~the permutation $(12)$, the Dessin~$D_1$ corresponding to the correlator
~$\langle\langle\mathcal{\tilde{R}}_{(id,id,id)}^{(1,1,1,1),(12)}\rangle\rangle_{I,\sharp_L=2}$ is determined by~
$(id,(12),(12))$ (see Fig. \ref{id12121}),
Let us delete white vertices in the Dessin assigned to the operator~$\mathcal{\tilde{R}}_{(id,id,id)}^{(1,1,1,1),(12)}$
and black vertices in the Dessin~$D_1$ corresponding to the correlator
$\langle\langle\mathcal{\tilde{R}}_{(id,id,id)}^{(1,1,1,1),(12)}\rangle\rangle_{I,\sharp_L=2}$,
and glue the remaining vertices with the valence kept.
It gives the Dessin~$D_2$ corresponding to the correlator
~$\langle\langle\mathcal{\tilde{R}}_{(id,id,id)}^{(1,1,1,1),(12)}\rangle\rangle_{I,\sharp_L=2}$
(see Fig.\ref{id12122}).

\begin{figure}[H]
\centering
\includegraphics[height=3cm]{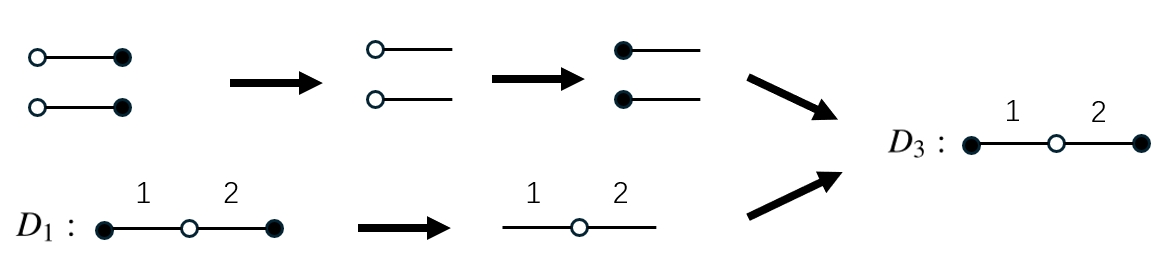}
\caption{The Dessin~$D_3$ of $\langle\langle\mathcal{\tilde{R}}_{(id,id,id)}^{(1,1,1,1),(12)}\rangle\rangle_{I,\sharp_L=2}$.}
\label{id12123}
\end{figure}

Let us delete black vertices in the Dessin assigned to the operator $\mathcal{\tilde{R}}_{(id,id,id)}^{(1,1,1,1),(12)}$
and convert the remaining white vertices to black vertices. Then we delete black vertices in the Dessin~$D_1$ corresponding
to the correlator $\langle\langle\mathcal{\tilde{R}}_{(id,id,id)}^{(1,1,1,1),(12)}\rangle\rangle_{I,\sharp_L=2}$ again,
and glue the remaining vertices with the valence kept. It gives the Dessin~$D_3$ corresponding to the correlator
~$\langle\langle\mathcal{\tilde{R}}_{(id,id,id)}^{(1,1,1,1),(12)}\rangle\rangle_{I,\sharp_L=2}$ (see Fig.\ref{id12123}).
It is obvious that~$\chi(D_1)=\chi(D_2)=\chi(D_3)=2$ from Figs. \ref{id12121}-\ref{id12123} and~$\mathcal{N}_1=1$.
Thus from the second line of~(\ref{twotensor}), we reach the result of the last line of (\ref{correlators}).

When~$\sharp_L=1$, the rainbow model (\ref{rainbowr3}) becomes the Aristotelian model~\cite{ItoyamaJHEP2017}.
The extreme case is that of $\sharp_L=1$ and $t_{\tilde{\sigma}}^{(a_1,0,\cdots,0,b_1,0,\cdots,0)}=0$ unless
$\sigma_1=(1,\cdots,a_1),\sigma_2=\sigma_{3}=id$ in (\ref{rainbowr3}).
It gives the red rainbow model \cite{Mishnyakov}
\begin{eqnarray}\label{rainbowred}
\mathcal{Z}_{red}&=&\frac{1}{\tilde{\mathcal{Z}}_{red}}
\int dTd\bar{T}\exp(-Tr T\bar{T}
+\sum_{a_1=1}^{\infty}\sum_{level (\mathcal{R}_{\tilde{\sigma}}^{(a_1,0,\cdots,0,b_1,0,\cdots,0)})=a_1}
t_{\tilde{\sigma}}^{(a_1,0,\cdots,0,b_1,0,\cdots,0)}
\mathcal{R}_{\tilde{\sigma}}^{(a_1,0,\cdots,0,b_1,0,\cdots,0)})\nonumber\\
&=&\sum_{R}\frac{D_R(N_1)D_R(N_2N_3)}{d_R}\chi_R(t),
\end{eqnarray}
where $R$ is the Young diagram, $D_R(N)$ is the dimension of representation~$R$ of the Lie algebra~$GL(N)$ and $\chi_R(t)$ is the Schur polynomial,
and $\tilde{\mathcal{Z}}_{red}=\int dT d\bar{T}\exp(-\Tr T\bar{T})$.

The correlators are
\begin{eqnarray}\label{redcorre}
\langle\langle  \prod_{i=1}^{l}\mathcal{R}_{\tilde{\sigma}_i}
^{(a_1^{(i)},0,\cdots,0,b_1^{(i)},0,\cdots,0)}\rangle\rangle_{I,\sharp_L=1}
&=&\frac{l!}{k!\lambda_{(\tilde{\sigma}_1,\cdots,\tilde{\sigma}_l)}}
\sum_{\tau}{P}_{~\tau(\tilde{\sigma}_1),\cdots,\tau(\tilde{\sigma}_l)}
^{(a_1^{(1)},0,\cdots,0,b_1^{(1)},0,\cdots,0),\cdots,
(a_1^{(l)},0,\cdots,0,b_1^{(l)},0,\cdots,0)}\nonumber\\
&=&\frac{1}{\mathcal{N}_1}\sum_{D_1,D_2,D_3\in{D_{col}}}
\frac{N_1^{\frac{1}{2}\chi(D_1)}N_2^{\chi(D_2)-\frac{1}{2}\chi(D_1)}N_3^{\chi(D_3)
-\frac{1}{2}\chi(D_1)}}{|Aut(D_{1})|\cdot |Aut(D_{2})|\cdot |Aut(D_{3})|}\nonumber\\
&=&\sum_{R\vdash \sum_{j=1}^{l}a_1^{(j)}}\frac{D_R(N_1)D_R(N_2N_3)}{d_R}\psi_R(\sigma_1^{-1}),
\end{eqnarray}
where $\psi_R(\sigma_1^{-1})$ is the character of the symmetric group
$S_{|R|}$.~$d_R=\frac{\psi_R(id)}{(a_1^{(1)}+\cdots+a_1^{(l)})!}$.

\begin{figure}[H]
\centering
\includegraphics[height=3cm]{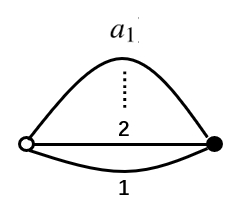}
\caption{Dessins corresponding to~$\mathcal{R}_{\tilde{\sigma}}^{(a_1,0,\cdots,0,b_1,0,\cdots,0)}$.}
\label{redmodel}
\end{figure}

The operators~$\mathcal{R}_{\tilde{\sigma}}^{(a_1,0,\cdots,0,b_1,0,\cdots,0)}$ are assigned to the Dessins in Fig. \ref{redmodel},
The Dessins~$D_1$ assigned to correlators
$ \langle\langle\mathcal{R}_{\tilde{\sigma}}^{(a_1,0,\cdots,0,b_1,0,\cdots,0)} \rangle\rangle_{I,\sharp_L=1}$
are obtained by taking any permutation in~$S_{a_1}$ to label the edges and going around the white vertices.
The Dessins~$D_2$ and~$D_3$ are the same. There is only one black vertex in them, which is connected by any possible
white vertices.

\subsection{Another rainbow tensor model}
By means of the connected operators $\mathcal{R}_{\tilde{\sigma}}^{(\vec{a},\vec{b})}$,
we can construct another rainbow tensor model with rank-$3$
\begin{eqnarray}\label{inmodel}
\mathcal{Z}_{II}&=&
\frac{1}{\tilde{\mathcal{Z}}_{II}}\int \prod_{j=1}^{\sharp_L}dT^j d\bar{T}_j\exp(-\sum_{j=1}^{\sharp_L-1}\Tr T^j\bar{T}_{j+1}-\Tr T^{\sharp_L}\bar{T}_{1}
+\sum_{a=1}^{\infty}\sum_{level (\mathcal{R}_{\tilde{\sigma}}^{(\vec{a},\vec{b})})=a}
t_{\tilde{\sigma}}^{(\vec{a},\vec{b})}\mathcal{R}_{\tilde{\sigma}}^{(\vec{a},\vec{b})})\nonumber\\
&=&\exp(\mathcal{\hat{W}}_{II})\cdot 1,
\end{eqnarray}
where
\begin{eqnarray}
\tilde{\mathcal{Z}}_{II}&=&\int \prod_{j=1}^{\sharp_L}dT^j d\bar{T}_j\exp(-\sum_{j=1}^{\sharp_L-1}\Tr T^j\bar{T}_{j+1}-\Tr T^{\sharp_L}\bar{T}_{1}).
\end{eqnarray}
Here the $W$-representation proceeds similarly with the case of (\ref{rainbowr3}). The operator $\mathcal{\hat{W}}_{II}$ is given by
\begin{eqnarray}\label{inrwope}
\mathcal{\hat{W}}_{II}
&=&\sum_{c=1}^{\infty}\dsum_{level (\mathcal{R}_{\tilde{\alpha}}^{(\vec{c},\vec{d})})=c}
\sum_{k=1}^3\sum_{j=1}^{\sharp_L}\sum_{\Gamma_5}
\sum_{\tilde{\beta}_1,\cdots,\tilde{\beta}_k}(1-\delta_{c_j,1})
(\tilde{\Delta}_j)_{(\tilde{\alpha},\tilde{\beta}_1,\cdots,\tilde{\beta}_k)}^{\bar\nu}
t_{\tilde{\alpha}}^{(\vec{c},\vec{d})}
\frac{\partial}{\partial t_{\tilde{\beta}_1}^{(\vec{c}^{(1)},\vec{d}^{(1)})}}\cdots
\frac{\partial}{\partial t_{\tilde{\beta}_k}^{(\vec{c}^{(k)},\vec{d}^{(k)})}}
\nonumber\\
&+&\sum_{a,c=1}^{\infty}\dsum_{\Upsilon_8}
\dsum_{\Upsilon_9}
\sum_{j=1}^{\sharp_L}{(\tilde{\Lambda}_j)}_{(\tilde{\sigma},\tilde{\alpha},\tilde{\beta})}
^{((\vec{a},\vec{b}),(\vec{c},\vec{d}),\bar\xi_4)}
t_{\tilde{\alpha}}^{(\vec{c},\vec{d})} t_{\tilde{\sigma}}^{(\vec{a},\vec{b})}
\frac{\partial}{\partial t_{\tilde{\beta}}^{\bar\xi_4}}
+\sum_{j=1}^{\sharp_L}t_{(id,id,id)}^{(\overbrace{\scriptstyle 0,\cdots,0}^{j-1},1,
\overbrace{\scriptstyle 0,\cdots,0}^{\sharp_L},1,\overbrace{\scriptstyle 0,\cdots,0}^{\sharp_L-j-1})}N_1N_2N_3\nonumber\\
&+&t_{(id,id,id)}^{(\overbrace{\scriptstyle 0,\cdots,0}^{\sharp_L-1},1,1
\overbrace{\scriptstyle 0,\cdots,0}^{\sharp_L-1})}N_1N_2N_3,
\end{eqnarray}
in which $ \Gamma_5= \{\sum_{i=1}^k c^{(i)}_j+1=c_j, \sum_{i=1}^kd^{(i)}_{j+1}+1=d_{j+1}, \sum_{i=1}^kc^{(i)}_m=c_m,
\sum_{i=1}^kd^{(i)}_n=d_n,m\neq j, n\neq j+1,c_1^{(1)}\leq \cdots\leq c_1^{(k)}\}$,
$\Upsilon_8=\{level (\mathcal{R}_{\tilde{\sigma}}^{(\vec{a},\vec{b})})=a,level (\mathcal{R}_{\tilde{\alpha}}^{(\vec{c},\vec{d})})=c\}$,
$\Upsilon_9=\{\tilde{\beta},level(\mathcal{R}_{\tilde{\beta}}^{\bar\xi_4})
	=level (\mathcal{R}_{\tilde{\sigma}}^{(\vec{a},\vec{b})})+level (\mathcal{R}_{\tilde{\alpha}}^{(\vec{c},\vec{d})})-1\}$
and $\bar\xi_4=(\vec{a}+\vec{c}-\vec{(1)}_j,\vec{b}+\vec{d}-\vec{(1)}_{j+1})$.

In (\ref{inrwope}), $(\tilde{\Delta}_j)_{(\tilde{\alpha},\tilde{\beta}_1,\cdots,\tilde{\beta}_k)}^{\bar\nu},j=1,\cdots,\sharp_L$,
are polynomials of $N_i$ with integer coefficients given by cut operations on~$\mathcal{R}_{\tilde{\alpha}}^{(\vec{a},\vec{b})}$
\begin{eqnarray}\label{incut}
\Delta_{II,j} \mathcal{R}_{\tilde{\alpha}}^{(\vec{a},\vec{b})}
&\equiv&\sum_{i_1=1}^{N_{1}}\sum_{i_2=1}^{N_{2}}
\dfrac{\partial^2 \mathcal{R}_{\tilde{\alpha}}^{(\vec{a},\vec{b})}
}{\partial T^j_{i_1,i_2}
\partial \bar{T}_{j+1}^{i_1,i_2}}\nonumber\\
&=&\sum_{k=1}^3\sum_{\Gamma_6}
\sum_{\tilde{\beta}_1,\cdots,\tilde{\beta}_k}
(\tilde{\Delta}_j)_{(\tilde{\alpha},\tilde{\beta}_1,\cdots,\tilde{\beta}_k)}
^{\bar\mu}\
\mathcal{R}_{\tilde{\beta}_1}^{(\vec{a}^{(1)},\vec{b}^{(1)})}\cdots
\mathcal{R}_{\tilde{\beta}_k}^{(\vec{a}^{(k)},\vec{b}^{(k)})},
\end{eqnarray}
where we set $j+1$ to be $\sharp_L+1=1$ for $j=\sharp_L$,  $ level(\mathcal{R}_{\tilde{\alpha}}^{(\vec{a},\vec{b})})\geqslant 2$,
and $\Gamma_6=\{ \sum_{i=1}^ka^{(i)}_j+1=a_j, \sum_{i=1}^kb^{(i)}_{j+1}+1=b_{j+1}, \sum_{i=1}^ka^{(i)}_m=a_m,
\sum_{i=1}^kb^{(i)}_n=b_n,m\neq j, n\neq j+1,a_1^{(1)}\leq \cdots\leq a_1^{(k)}\}$.

The integer coefficients $(\tilde{\Lambda}_j)_{(\tilde{\sigma},\tilde{\alpha},\tilde{\beta})} ^{((\vec{a},\vec{b}),(\vec{c},\vec{d}),
\bar\xi_4)}$ in (\ref{inrwope}) are given by join operations on
$\mathcal{R}_{\tilde{\sigma}}^{(\vec{a},\vec{b})}$ and $\mathcal{R}_{\tilde{\alpha}}^{(\vec{c},\vec{d})}$
\begin{eqnarray}\label{joininm}
&&\{\mathcal{R}_{\tilde{\sigma}}^{(\vec{a},\vec{b})},
\mathcal{R}_{\tilde{\alpha}}^{(\vec{c},\vec{d})}\}_{II,j}
=
\sum_{i_1=1}^{N_{1}}\sum_{i_2=1}^{N_{2}}
\dfrac{\partial \mathcal{R}_{\tilde{\sigma}}^{(\vec{a},\vec{b})}}{\partial T^j_{i_1,i_2}}
\dfrac{\partial \mathcal{R}_{\tilde{\alpha}}^{(\vec{c},\vec{d})}}{\partial \bar{T}_{j+1}^{i_1,i_2}}
=\sum_{\Upsilon_9}
(\tilde{\Lambda}_j)_{(\tilde{\sigma},\tilde{\alpha},\tilde{\beta})}
^{((\vec{a},\vec{b}),(\vec{c},\vec{d}), \bar\xi_4)}
\mathcal{R}_{\tilde{\beta}}^{\bar\xi_4}.
\end{eqnarray}

The correlators are given by
\begin{eqnarray}
\langle \langle \prod_{i=1}^{l}\mathcal{R}_{\tilde{\sigma}_i}^{(\vec{a}^{(i)},\vec{b}^{(i)})}\rangle\rangle_{II,\sharp_L}
&\equiv&\frac{1}{\tilde{\mathcal{Z}}_{II}}\int \dprod_{j=1}^{\sharp_L}dT^j d\bar{T}_j\mathcal{R}_{\tilde{\sigma}_1}^{(\vec{a}^{(1)},\vec{b}^{(1)})}
 \cdots\mathcal{R}_{\tilde{\sigma}_l}^{(\vec{a}^{(l)},\vec{b}^{(l)})}
 \exp(- \sum_{j=1}^{\sharp_L-1}\Tr T^j\bar{T}_{j+1}-\Tr T^{\sharp_L}\bar{T}_{1}) \nonumber\\
&=&\frac{l!}{k!\lambda_{(\tilde{\sigma}_1,\cdots,\tilde{\sigma}_l)}}
\sum_{\tau}{P}_{~\tau(\tilde{\sigma}_1),\cdots,\tau(\tilde{\sigma}_l)}
^{(\vec{a}^{(1)},\vec{b}^{(1)}),\cdots,
(\vec{a}^{(l)},\vec{b}^{(l)})},
\end{eqnarray}
where ${P}_{~\tau(\tilde{\sigma}_1),\cdots,\tau(\tilde{\sigma}_l)}
^{(\vec{a}^{(1)},\vec{b}^{(1)}),\cdots,
(\vec{a}^{(l)},\vec{b}^{(l)})}$ are coefficients of
$t_{\tau(\tilde{\sigma}_1)} \cdots \cdot t_{\tau(\tilde{\sigma}_l)}$
in the $k$-th power of the operator $\mathcal{\hat{W}}_{II}$, and~$k=\sum_{i=1}^{\sharp_L}\sum_{j=1}^{l}a_i^{(j)}$.
It is obvious that correlators are zero unless
$\sum_{j=1}^{l}a_i^{(j)}=\sum_{j=1}^{l}b_{i+1}^{(j)},i=1,\cdots,\sharp_L-1$, and $\sum_{j=1}^{l}a_{\sharp_L}^{(j)}=\sum_{j=1}^{l}b_{1}^{(j)}$.

In similarity with the case of (\ref{wickth}), we have
\begin{eqnarray}\label{wickth2}
&&\langle\langle\mathcal{R}_{\tilde{\sigma}}^{(a_1,\cdots,a_{\sharp_L},b_1,\cdots,b_{\sharp_L})}\rangle\rangle_{II,\sharp_L}
=\langle\langle\mathcal{\tilde{R}}_{(id,\sigma_2\circ\sigma_1^{-1},\sigma_3\circ\sigma_1^{-1})}
^{(a_1,\cdots,a_{\sharp_L},b_1,\cdots,b_{\sharp_L}),\sigma_1}\rangle\rangle_{II,\sharp_L}\nonumber\\
&=&\sum_{\Theta_3}
\sum_{\tau_1,\tau_2,\tau_3\in S_{a_1+ \cdots+a_{\sharp_L}}}
\delta(\sigma_1\circ\gamma\circ\tau_1)\delta(\sigma_2\circ\gamma\circ\tau_2)\delta(\sigma_3\circ\gamma\circ\tau_3)
N_1^{\sharp(\tau_1)}N_2^{\sharp(\tau_2)}N_3^{\sharp(\tau_3)}\nonumber\\
&=&\frac{1}{\mathcal{N}_2}
 \sum_{\Theta_4}
\sum_{\Theta_5}
\sum_{\tau_1,\tau_2,\tau_3\in S_{a_1+\cdots+a_{\sharp_L}}} \frac{1}{(a_1!\cdots a_{\sharp_L}!)^3}
\delta(\gamma\circ\tau_1)\delta(\sigma_2\circ\sigma_1^{-1}\circ\gamma\circ\tau_2)
\delta(\sigma_3\circ\sigma_1^{-1}\circ\gamma\circ\tau_3)\nonumber\\
&&
\cdot
N_1^{\sharp(\gamma)}
N_2^{\sharp(\tau_2)+\sharp{(\sigma_2}\circ\sigma_1^{-1})-\sum_{i=1}^{\sharp_L}a_i}
N_3^{\sharp(\tau_3)+\sharp{(\sigma_3}\circ\sigma_1^{-1})-\sum_{i=1}^{\sharp_L}a_i}\nonumber\\
&=&\frac{1}{\mathcal{N}_2}
 \sum_{D_1,D_2,D_3\in{D_{col}}}\frac{N_1^{\frac{1}{2}\chi(D_1)} N_2^{\chi(D_2)
 -\frac{1}{2}\chi(D_1)}N_3^{\chi(D_3)-\frac{1}{2}\chi(D_1)}}{|Aut(D_{1})|\cdot |Aut(D_{2})|\cdot |Aut(D_{3})|},
\end{eqnarray}
where $\Theta_3=\{\gamma, \gamma\circ\varphi \in S_{a_{\sharp_L}}\times S_{a_1}\times \cdots\times S_{a_{\sharp_L-1}}\}$,
$\Theta_4=\{ \sigma_2\in [\sigma_2]_{a_{\sharp_L},a_1,\cdots,a_{\sharp_L-1}},
\sigma_3\in [\sigma_3]_{a_{\sharp_L},a_1,\cdots,a_{\sharp_L-1}}\}$,
$\Theta_5=\{\gamma,\sigma_1^{-1}\circ\gamma\circ\varphi\in S_{a_{\sharp_L}}\times S_{a_1}\times \cdots\times S_{a_{\sharp_L-1}}\}$,
 $\varphi(i)=\sum_{j=1}^{\sharp_L-1}a_j+i,i=1,\cdots,a_{\sharp_L},\varphi(i+a_{\sharp_L})=i,i=1,\cdots,\sum_{j=1}^{\sharp_L-1}a_j$,
and $[\alpha]_{a_{\sharp_L},a_1,\cdots,a_{\sharp_L-1}}$ is the set of permutations in $S_{a_1+\cdots+a_{\sharp_L}}$
related to $\alpha$ by conjugation with elements in~$S_{a_{\sharp_L}}\times S_{a_1}\times \cdots\times S_{a_{\sharp_L-1}}$,
$\mathcal{N}_2=\frac{ 1}{(a_1!\cdots a_{\sharp_L}!)^3}\mid[\sigma_2]_{a_{\sharp_L},a_1,\cdots,a_{\sharp_L-1}}\mid
\cdot \mid [\sigma_3]_{a_{\sharp_L},a_1,\cdots,a_{\sharp_L-1}}\mid \cdot N_2^{\sharp{(\sigma_2\circ\sigma_1^{-1})}-\sum_{i=1}^{\sharp_L}a_i}
N_3^{\sharp{(\sigma_3\circ\sigma_1^{-1})}-\sum_{i=1}^{\sharp_L}a_i}$.

Here a permutation~$\gamma$ in the third line of (\ref{wickth2}) gives a coloring of the cyclically ordered edges coming out of a white
vertex of three Dessins~$D_1,D_2,D_3$. The Dessins~$D_1,D_2,D_3$ are obtained in the same way as in (\ref{wickth}).
There are~$\sharp_L$ colors in colored Dessins~$D_1,D_2,D_3$, which represent~$\sharp_L$ types of contractions between
$T^{\sharp_L},T^{1},\cdots,T^{\sharp_L-1}$ and $\bar{T}_{1},\cdots,\bar{T}_{\sharp_L}$, respectively.

When particularized to the case of two tensors~$T^1$ and~$T^2$ in (\ref{inmodel}), it gives
\begin{eqnarray}\label{rainbow23}
\mathcal{Z}_{II,2T}&=&\frac{1}{\tilde{\mathcal{Z}}_{II,2T}}
\int \prod_{j=1}^{2}dT^j d\bar{T}_j\exp(-\Tr T^1\bar{T}_2
+\sum_{a_1+a_2=1}^{\infty}\sum_{level (\mathcal{R}_{\tilde{\sigma}}^{(a_1,a_{2},b_1,b_{2})})=a_1+a_2}
t_{\tilde{\sigma}}^{(a_1,a_{2},b_1,b_{2})}\mathcal{R}_{\tilde{\sigma}}^{(a_1,a_{2},b_1,b_{2})}\nonumber\\
&&-\Tr T^2\bar{T}_1)
\nonumber\\
&=&\exp(\mathcal{\hat{W}}'_{II})\cdot 1,
\end{eqnarray}
where
\begin{eqnarray}
\tilde{\mathcal{Z}}_{II,2T}&=&\int \prod_{j=1}^{2}dT^j d\bar{T}_j\exp(-\Tr T^1\bar{T}_2-\Tr T^2\bar{T}_1),
\end{eqnarray}
and
\begin{eqnarray}\label{operatorin23}
\mathcal{\hat{W}}'_{II}
&=&\sum_{\bar\eta=1}^{\infty}\dsum_{level (\mathcal{R}_{\tilde{\alpha}}^{\bar\eta})=\bar\eta}
\sum_{k=1}^3\sum_{\Gamma_7}
\sum_{\tilde{\beta}_1,\cdots,\tilde{\beta}_k}(1-\delta_{c_1,1})
(\tilde{\Delta}_1)_{(\tilde{\alpha},\tilde{\beta}_1,\cdots,\tilde{\beta}_k)}
^{(\bar\eta,\bar\eta_1,\cdots,\bar\eta_k)}t_{\tilde{\alpha}}^{\bar\eta}
\frac{\partial}{\partial t_{\tilde{\beta}_1}^{\bar\eta_1}}\cdots
\frac{\partial}{\partial t_{\tilde{\beta}_k}^{\bar\eta_k}}
\nonumber\\
&&
\nonumber\\
&+&\sum_{\bar\eta=1}^{\infty}\dsum_{level (\mathcal{R}_{\tilde{\alpha}}^{\bar\eta})=\bar\eta}
\sum_{k=1}^3\sum_{\Gamma_8}
\sum_{\tilde{\beta}_1,\cdots,\tilde{\beta}_k}(1-\delta_{c_2,1})
(\tilde{\Delta}_2)_{(\tilde{\alpha},\tilde{\beta}_1,\cdots,\tilde{\beta}_k)}
^{(\bar\eta,\bar\eta_1,\cdots,\bar\eta_k)}t_{\tilde{\alpha}}^{\bar\eta}
\frac{\partial}
{\partial t_{\tilde{\beta}_1}^{\bar\eta_1}}\cdots
\frac{\partial}{\partial t_{\tilde{\beta}_k}^{\bar\eta_k}}
\nonumber\\
&&
\nonumber\\
&+&\sum_{\bar\rho,\bar\eta=1}^{\infty}\dsum_{\Upsilon_{10}}
\dsum_{\Upsilon_{11}}
{(\tilde{\Lambda}_1)}_{(\tilde{\sigma},\tilde{\alpha},\tilde{\beta})}
^{(\bar\rho,\bar\eta,\bar{\xi}_5)}
t_{\tilde{\sigma}}^{\bar\rho}t_{\tilde{\alpha}}^{\bar\eta}
\frac{\partial}{\partial t_{\tilde{\beta}}^{\bar{\xi}_5}}
+\sum_{\bar\rho,\bar\eta=1}^{\infty}\dsum_{\Upsilon_{10}}
\dsum_{\Upsilon_{12}}
{(\tilde{\Lambda}_2)}_{(\tilde{\sigma},\tilde{\alpha},\tilde{\beta})}
^{(\bar\rho,\bar\eta,\bar\xi_6)}
t_{\tilde{\sigma}}^{\bar\rho}t_{\tilde{\alpha}}^{\bar\eta}
\frac{\partial}{\partial t_{\tilde{\beta}}^{\bar\xi_6}}
\nonumber\\&
+&t_{(id,id,id)}^{(1,0,0,1)}N_1N_2N_3+t_{(id,id,id)}^{(0,1,1,0)}N_1N_2N_3,
\end{eqnarray}
in which~$\Gamma_7=\{ \sum_{i=1}^kc^{(i)}_1+1=c_2, \sum_{i=1}^kd^{(i)}_{1}+1=d_{2},
\sum_{i=1}^kc^{(i)}_2=c_1,\sum_{i=1}^kd^{(i)}_2=d_1, c_1^{(1)}\leq \cdots\leq c_1^{(k)}\}$,
$\Gamma_8= \{\sum_{i=1}^kc^{(i)}_2+1=c_1, \sum_{i=1}^kd^{(i)}_{2}+1=d_{1},
\sum_{i=1}^kc^{(i)}_1=c_2,\sum_{i=1}^kd^{(i)}_1=d_2, c_1^{(1)}\leq \cdots\leq c_1^{(k)}\}$,
$\Upsilon_{10}=\{level (\mathcal{R}_{\tilde{\sigma}}^{\bar\rho})=\bar\rho,level (\mathcal{R}_{\tilde{\alpha}}^{\bar\eta})=\bar\eta\}$,
$\Upsilon_{11}=\{\tilde{\beta},level(\mathcal{R}_{\tilde{\beta}}^{\bar{\xi}_5})
=level (\mathcal{R}_{\tilde{\sigma}}^{\bar\rho})+ level (\mathcal{R}_{\tilde{\alpha}}^{\bar\eta})-1\}$,
$\Upsilon_{12}=\{\tilde{\beta},level(\mathcal{R}_{\tilde{\beta}}^{\bar{\xi}_6})
=level (\mathcal{R}_{\tilde{\sigma}}^{\bar\rho})+ level (\mathcal{R}_{\tilde{\alpha}}^{\bar\eta})-1\}$,
$\bar{\xi}_5=(a_1+c_1-1,a_2+c_2,b_1+d_1,b_{2}+d_{2}-1)$ and
$\bar\xi_6=(a_1+c_1,a_2+c_2-1,b_1+d_1-1,b_{2}+d_{2})$.

The correlators are given by
\begin{eqnarray}\label{intertensor}
\langle \langle \prod_{i=1}^{l}\mathcal{R}_{\tilde{\sigma}_i}
^{(a_1^{(i)},a_{2}^{(i)},b_1^{(i)},b_{2}^{(i)})}\rangle\rangle_{II,\sharp_L=2}
&=&\frac{l!}{k!\lambda_{(\tilde{\sigma}_1,\cdots,\tilde{\sigma}_l)}}
\sum_{\tau}{P}_{~\tau(\tilde{\sigma}_1),\cdots,\tau(\tilde{\sigma}_l)}
^{(a_1^{(1)},a_{2}^{(1)},b_1^{(1)},b_{2}^{(1)}),\cdots,
(a_1^{(l)},a_{2}^{(l)},b_1^{(l)},b_{2}^{(l)})}\nonumber\\
&=&\frac{1}{\mathcal{N}_2}{\sum_{D_1,D_2,D_3\in{D_{col}}}
\frac{N_1^{\frac{1}{2}\chi(D_1)}N_2^{\chi(D_2)-\frac{1}{2}\chi(D_1)}
N_3^{\chi(D_3)-\frac{1}{2}\chi(D_1)}}{|Aut(D_{1})|\cdot |Aut(D_{2})|\cdot |Aut(D_{3})|}}.
\end{eqnarray}

Let us list some correlators
\begin{eqnarray}\label{correlatorsinten}
 \langle\langle \mathcal{R}_{(id,id,id)}^{(1,1,1,1)}\rangle\rangle_{II,\sharp_L=2} &=&N_1N_2N_3,\nonumber\\
 \langle\langle \mathcal{R}_{(id,(12),(12))}^{(1,1,1,1)}\rangle\rangle_{II,\sharp_L=2} &=&N_1N_2^2N_3^2,\nonumber\\
 \langle \langle\mathcal{R}_{((12),id,(12))}^{(1,1,1,1)}\rangle\rangle_{II,\sharp_L=2} &=&N_1^2N_2N_3^2,\nonumber\\
 \langle \langle\mathcal{R}_{((12),(12),id)}^{(1,1,1,1)}\rangle\rangle_{II,\sharp_L=2} &=&N_1^2N_2^2N_3,\nonumber\\
 \langle \langle\mathcal{R}_{(id,id,(12))}^{(1,1,1,1)}\rangle\rangle_{II,\sharp_L=2} &=&N_1N_2N_3^2,\nonumber\\
 \langle \langle\mathcal{R}_{(id,(12),id)}^{(1,1,1,1)}\rangle\rangle_{II,\sharp_L=2} &=&N_1N_2^2N_3,\nonumber\\
 \langle \langle\mathcal{R}_{((12),id,id)}^{(1,1,1,1)}\rangle\rangle_{II,\sharp_L=2} &=&N_1^2N_2N_3,\nonumber\\
 \langle \langle\mathcal{R}_{((12),(12),(12))}^{(1,1,1,1)}\rangle\rangle_{II,\sharp_L=2} &=&N_1^2N_2^2N_3^2.
 \end{eqnarray}

\section{Complex multi-matrix models}

Let $t_{\tilde{\sigma}}^{(\vec{a},\vec{b})}
=\frac{1}{N_{(\sigma_2,\sigma_3)}}t_{(\sigma_2,\sigma_3)}^{(\vec{a},\vec{b})}
=\frac{1}{N_{(\sigma_2,\sigma_3)}}t_{(\breve{\sigma})}^{(\vec{a},\vec{b})}$
and $N_1=1$ in (\ref{rainbowr3}) and (\ref{inmodel}),
where~$N_{(\sigma_2,\sigma_3)}$ denotes the number of operators with the last two permutations
in~$\tilde{\sigma}$ equaled~$\sigma_2$ and $\sigma_3$,
the rainbow models reduce to the following matrix models:

\begin{eqnarray}\label{matrixmul}
Z_{I}&=&\frac{1}{\tilde{Z}_{I}}
\int \prod_{j=1}^{\sharp_L}dM^j d\bar{M}_j\exp(
\sum_{a=1}^{\infty}\sum_{level (\mathcal{M}_{(\breve{\sigma})})=a}
t_{(\breve{\sigma})}^{(\vec{a},\vec{b})}
\mathcal{M}_{(\breve{\sigma})}^{(\vec{a},\vec{b})}-\sum_{j=1}^{\sharp_L}\Tr M^j\bar{M}_j)\nonumber\\
&=&\exp(\hat{W}_{I})\cdot 1,
\end{eqnarray}
and
\begin{eqnarray}\label{inmmodl}
Z_{II}&=&\frac{1}{\tilde{Z}_{II}}
\int \prod_{j=1}^{\sharp_L}dM^j d\bar{M}_j\exp(
\sum_{a=1}^{\infty}\sum_{level (\mathcal{M}_{(\breve{\sigma})})=a}
t_{(\breve{\sigma})}^{(\vec{a},\vec{b})}
\mathcal{M}_{(\breve{\sigma})}^{(\vec{a},\vec{b})}-\sum_{j=1}^{\sharp_L-1}\Tr M^j\bar{M}_{j+1}-\Tr M^{\sharp_L}\bar{M}_{1})\nonumber\\
&=&\exp(\hat{W}_{II})\cdot 1,
\end{eqnarray}
where
\begin{eqnarray}
\tilde{Z}_{I}&=&\int \prod_{j=1}^{\sharp_L}dM^j d\bar{M}_j\exp(-\sum_{j=1}^{\sharp_L}\Tr M^j\bar{M}_j),\nonumber\\
\tilde{Z}_{II}&=&\int \prod_{j=1}^{\sharp_L}dM^j d\bar{M}_j\exp(-\sum_{j=1}^{\sharp_L-1}\Tr M^j\bar{M}_{j+1}-\Tr M^{\sharp_L}\bar{M}_{1}),
\end{eqnarray}
the $W$-operators are given by
$\hat{W}_{I}=\mathcal{\hat{W}}_{I}\big|_{t_{\tilde{\sigma}}^{(\vec{a},\vec{b})}=\frac{1}{N_{(\sigma_2,\sigma_3)}}
t_{(\breve{\sigma})}^{(\vec{a},\vec{b})},N_1=1}$
and
$\hat{W}_{II}=$
$\mathcal{\hat{W}}_{II}\big|_{t_{\tilde{\sigma}}^{(\vec{a},\vec{b})}=\frac{1}{N_{(\sigma_2,\sigma_3)}}
t_{(\breve{\sigma})}^{(\vec{a},\vec{b})},N_1=1}$,
and
\begin{eqnarray}\label{}
\mathcal{M}_{(\breve{\sigma})}^{(\vec{a},\vec{b})}=
\mathcal{M}_{(\sigma_2,\sigma_3)}^{(\vec{a},\vec{b})}
=\mathcal{M}_{I_2,I_3}\cdot(\sigma_2,\sigma_3)\cdot\bar{\mathcal{M}}^{I_2,I_3}
=\mathcal{M}_{I_2,I_3}\bar{\mathcal{M}}^{\sigma_2(I_2),\sigma_3(I_3)},
\end{eqnarray}
\begin{eqnarray}
\mathcal{M}_{I_2,I_3}&=&M^1_{i_2^{(1)},i^{(1)}_3}\cdots M^1_{i_2^{(a_1)},i^{(a_1)}_3}\cdots
M^{\sharp_L}_{i_2^{(a_1+\cdots+a_{\sharp_L-1}+1)},i^{(a_1+\cdots+a_{\sharp_L-1}+1)}_3}\cdots
M^{\sharp_L}_{i_2^{(a_1+\cdots+a_{\sharp_L})},i^{(a_1+\cdots+a_{\sharp_L})}_3},\nonumber\\
\bar{\mathcal{M}}^{I_2,I_3}&=&\bar{M}_1^{i_2^{(1)},i^{(1)}_3}\cdots \bar{M}_1^{i_2^{(b_1)},i^{(b_1)}_3}
\cdots \bar{M}_{\sharp_L}^{i_2^{(b_1+\cdots+b_{\sharp_L-1}+1)},i^{(b_1+\cdots+b_{\sharp_L-1}+1)}_3}\cdots
\bar{M}_{\sharp_L}^{i_2^{(b_1+\cdots+b_{\sharp_L})},i^{(b_1+\cdots+b_{\sharp_L})}_3},\nonumber\\
\bar{\mathcal{M}}^{\sigma_2(I_2),\sigma_3(I_3)}&=&(\sigma_2,\sigma_3)\cdot\bar{\mathcal{M}}^{I_2,I_3}
=\bar{M}_1^{i_2^{\sigma_2(1)},i^{\sigma_3(1)}_3}\cdots\bar{M}_1^{i_2^{\sigma_2(b_1)},i^{\sigma_3(b_1)}_3}\cdots\nonumber\\
&&\bar{M}_{\sharp_L}^{i_2^{\sigma_2(b_1+\cdots+b_{\sharp_L-1}+1)},i^{\sigma_3(b_1+\cdots+b_{\sharp_L-1}+1)}_3}
\cdots \bar{M}_{\sharp_L}^{i_2^{\sigma_2(b_1+\cdots+b_{\sharp_L})},i_3^{\sigma_3(b_1+\cdots+b_{\sharp_L})}},
\end{eqnarray}
$M_{i_{2} ,i_3}^{j},j=1,\cdots,\sharp_L$, are complex matrices, $\bar{M}^{i_{2},i_{3}}_{j}$ are their complex conjugates
and $level (\mathcal{M}_{(\breve{\sigma})})=\sum_{i=1}^{\sharp_L}a_i$.

The correlators in the matrix model (\ref{matrixmul}) are
\begin{eqnarray}
\langle \mathcal{M}_{(\sigma_2,\sigma_3)}^{(\vec{a},\vec{b})} \rangle_{I}
&\equiv&\frac{1}{\tilde{Z}_{I}}\int \prod_{j=1}^{\sharp_L}dM^j d\bar{M}_j
\mathcal{M}_{(\sigma_2,\sigma_3)}^{(\vec{a},\vec{b})} \exp(- \sum_{j=1}^{\sharp_L}\Tr M^j\bar{M}_j)\nonumber\\
&=&\lim_{N_1\rightarrow\infty}\frac{1}{N_1^{a_1+\cdots+a_{\sharp_L}}}\sum_{\sigma_1\in S_{a_1}\times\cdots\times S_{a_{\sharp_L}}}
\langle\langle\mathcal{R}_{(\sigma_1,\sigma_2,\sigma_3)}^{(\vec{a},\vec{b})}\rangle\rangle_{I,\sharp_L}\nonumber\\
&=&\langle \langle\mathcal{R}_{(\sigma_1,\sigma_2,\sigma_3)}^{(\vec{a},\vec{b})} \rangle\rangle_{I,\sharp_L}\big|_{N_1=1}.
\end{eqnarray}
Similarly,
\begin{eqnarray}
\langle \mathcal{M}_{(\sigma_2,\sigma_3)}^{(\vec{a},\vec{b})} \rangle_{II}
&\equiv&\frac{1}{\tilde{Z}_{II}}\int \prod_{j=1}^{\sharp_L}dM^j d\bar{M}_j
\mathcal{M}_{(\sigma_2,\sigma_3)}^{(\vec{a},\vec{b})} \exp(-\sum_{j=1}^{\sharp_L-1}\Tr M^j\bar{M}_{j+1}-\Tr M^{\sharp_L}\bar{M}_{1})\nonumber\\
&=&\lim_{N_1\rightarrow\infty}\frac{1}{N_1^{a_1+\cdots+a_{\sharp_L}}}
\sum_{\sigma_1\circ\varphi \in S_{a_{\sharp_L}}\times S_{a_1}\times \cdots\times S_{a_{\sharp_L-1}}}
\langle\langle\mathcal{R}_{(\sigma_1,\sigma_2,\sigma_3)}^{(\vec{a},\vec{b})}\rangle\rangle_{II,\sharp_L}\nonumber\\
&=&\langle \langle\mathcal{R}_{(\sigma_1,\sigma_2,\sigma_3)}^{(\vec{a},\vec{b})} \rangle\rangle_{II,\sharp_L}\big|_{N_1=1}.
\end{eqnarray}

Let us restrict~$\sigma_2$ and $\sigma_3$ to be in the conjugacy class of
~$\sigma_2=id,\sigma_3=(1,\cdots, k_1)(k_1+1, \cdots, k_1+k_2)
\cdots(k_1+\cdots +k_{u-1},\cdots ,k_1+\cdots +k_{u})$ in (\ref{matrixmul}),
and take $t_{(\sigma_2,\sigma_3)}^{(\vec{a},\vec{b})}=0$ for other $\sigma_2$ and $\sigma_3$,
it reduces to the complex multi-matrix model
\begin{eqnarray}\label{mc}
Z_{mc}&=&\frac{1}{\tilde{Z}_{I}}
\int \prod_{j=1}^{\sharp_L}dM^j d\bar{M}_j\exp[-\sum_{j=1}^{\sharp_L}\Tr M^j\bar{M}_j
+\sum_{j=1}^{\sharp_L}\sum_{k=0}^{\infty}t_{k}^{(j)}\Tr(M^j\bar{M}_j)^k\nonumber\\
&+&\sum_{u=2}^{\infty}\sum_{\substack{j_1,\cdots,j_u=1\\j_a\neq j_{a+1}}}^{\sharp_L}\sum_{k_1,\cdots,k_u=1}^{\infty}
t_{k_1,\cdots,k_u}^{(j_1,\cdots,j_u)}
\Tr(M^{j_1}\bar{M}_{j_1})^{k_1}(M^{j_2}\bar{M}_{j_2})^{k_2}\cdots (M^{j_u}\bar{M}_{j_u})^{k_u}],
\end{eqnarray}
where~$a=1,\cdots,u-1$, and~$M^j,j=1,\cdots,\sharp_L,$ are $N_2\times N_3$ complex matrices.
When particularized to the $N_2=N_3$ complex matrices in (\ref{mc}), it gives the model discussed in Ref.\cite{largen}.

\section{Conclusion}

We have analyzed the counting gauge-invariant operators and presented the formula (\ref{indeope}) of counting number of the independent operators at each level-$l$
in terms of Hurwitz numbers. A one-to-one correspondence between connected operators and colored Dessins was established.
In terms of the connected operators, we constructed two rainbow tensor models (\ref{rainbowr3}) and (\ref{inmodel}) with multi-tensors of rank-$3$,
where the Gaussian potential~$\sum_{j=1}^{\sharp_L}\Tr T^j\bar{T}_j$ and the interaction~$\sum_{j=1}^{\sharp_L-1}\Tr T^j\bar{T}_{j+1}+\Tr T^{\sharp_L}\bar{T}_{1}$
are contained in these two models, respectively.
The well-known red rainbow and Aristotelian models correspond to the particular cases of the rainbow tensor model (\ref{rainbowr3}).
It was shown that the constructed rainbow tensor models
can be realized by the $W$-representations. The correlators were calculated by means of the corresponding $W$-operators.
On the other hand, due to the one-to-one correspondence between connected operators and colored Dessins, we derived another compact expressions of
correlators from the viewpoint of colored Dessins. As the degradations of the rainbow tensor models,
two complex multi-matrix models (\ref{matrixmul}) and (\ref{inmmodl}) with
$W$-representations were obtained. Their correlators can be given by those ones in corresponding rainbow tensor models under certain limits or
restricting conditions.

\section *{Acknowledgments}
We are grateful to the referee of this paper for a series of valuable comments.
This work is supported by the National Natural Science Foundation of China (No. 12375004) and Natural Science Foundation of Henan Province (No. 242300420647).


\end{document}